\documentclass[aps,prx,citeautoscript,showpacs,showkeys,groupedaddress,superscriptaddress,twocolumn]{revtex4-2}

\usepackage[utf8x]{inputenc}
\usepackage{color}
\usepackage{bbm} 
\usepackage{amsfonts,amsmath,amssymb,stmaryrd}
\usepackage{graphicx}
\usepackage{subfigure}  % use for side-by-side figures
\usepackage{hyperref}
\usepackage{epsfig}
\usepackage{mathrsfs}
\usepackage{verbatim}
\usepackage{centernot}

\usepackage{hyperref}
\usepackage{epsfig}
\usepackage{mathrsfs}
\usepackage{verbatim}
\usepackage{centernot}
\usepackage{siunitx}
\usepackage{physics}
\usepackage{wasysym}
\usepackage{interval}
\intervalconfig{soft open fences}
\usepackage{mathtools}
\usepackage{amsthm}

\usepackage[normalem]{ulem}

\usepackage{array}

\usepackage{bm}	% bold in math mode
\renewcommand{\vec}[1]{\bm{#1}}

\newcommand{\inplanecorr}[1]{C\left(#1\right)}
\newcommand{\systemsize}{L}
\newcommand{\numparticles}{N}

\newcommand{\resubmark}{}

\begin{document}

%%%%%%%%%%%%%%%%%%%%%%%%%%%%%%%%%%%%%%%%%%%%%%%%%%%%%%
\title{{\resubmark Topological disordered phases} of Rydberg spin excitations in a honeycomb lattice induced
by density-dependent Peierls phases 
}
%%%%%%%%%%%%%%%%%%%%%%%%%%%%%%%%%%%%%%%%%%%%%%%%%%%%%%

\author{Simon Ohler}
\affiliation{Department of Physics and Research Center OPTIMAS, University of Kaiserslautern, D-67663 Kaiserslautern, Germany}
%\affiliation{Department of Physics and Research Center OPTIMAS, University of Kaiserslautern, Germany}
\author{Maximilian Kiefer-Emmanouilidis}
\affiliation{Department of Physics and Research Center OPTIMAS, University of Kaiserslautern, D-67663 Kaiserslautern, Germany}
\affiliation{German Research Centre for Artificial Intelligence, Embedded Intelligence, D-67663 Kaiserslautern, Germany}
%\affiliation{Institute of Electronic Structure and Laser, FORTH, GR-71110 Heraklion, Crete, Greece}
\author{Michael Fleischhauer}
\affiliation{Department of Physics and Research Center OPTIMAS, University of Kaiserslautern, D-67663 Kaiserslautern, Germany}

\begin{abstract}
    We show that the nonlinear transport of bosonic excitations in a two-dimensional honeycomb lattice  of  spin-orbit coupled Rydberg atoms gives rise to disordered quantum phases which are {\resubmark topological and may} be candidates for quantum spin liquids.
    As recently demonstrated in [Lienhard \textit{et al.} Phys. Rev. X, \textbf{10}, 021031 (2020)] the spin-orbit coupling breaks time-reversal and chiral symmetries and leads to a tunable density-dependent complex hopping of {\resubmark spin excitations which behave as hard-core bosons}.
    Using exact diagonalization (ED) we numerically investigate the phase diagram resulting from the competition between density-dependent and direct transport terms {\resubmark as well as density-density interactions}. In mean-field approximation there is a phase transition from a condensate to a 120$^\circ$ phase when the amplitude of the complex hopping exceeds that of the direct one. In the full model a new phase emerges close to the mean-field critical point as a result of quantum correlations induced by the density-dependence of the complex hopping. We show that {\resubmark without density-density interactions} this phase is a genuine disordered one, has large spin chirality and is characterized by a non-trivial many-body Chern number. The Chern number is found to be robust to disorder. ED simulations of small lattices with up to {\resubmark 30} lattice sites {\resubmark give indications for} a non-degenerate ground state with finite spin and collective gaps and thus to a bosonic integer-quantum Hall (BIQH) phase, protected by $U(1)$ symmetry.
    {\resubmark On the other hand, while staying finite, the many-body gap varies substantially when different twisted boundary conditions are applied, which points to a gapless phase.}
    For very strong negative nonlinear hopping amplitudes we find another disordered regime with vanishing spin gap. This phase also has a large spin chirality and could be a gapless spin-liquid but lies outside the parameter regime experimentally accessible in the Rydberg system. 
\end{abstract}

\pacs{123}

\date{\today}
\maketitle

%%%%%%%%%%%%%%%%%%%%%%%%%%%%%%%%%%%%%%%%%%%%%%%%%%%%%%
\section{Introduction}\label{sect:introduction}
%%%%%%%%%%%%%%%%%%%%%%%%%%%%%%%%%%%%%%%%%%%%%%%%%%%%%%

Over recent years, Rydberg atoms have become a versatile and robust platform to explore many-body quantum {spin} physics in the regime of strong correlations \cite{weimer2010rydberg,schauss2012observation,Bernien2017,Browaeys2020,Surace2020,scholl2021quantum} and for quantum information processing \cite{PhysRevLett.85.2208,PhysRevLett.87.037901,gaetan2009observation,urban2009observation,Saffman2010}. Using high principal quantum numbers, Rydberg-excited atoms have sizable interactions even at distances of several $\mu$m, while their lifetime is on the order of ms.
{\resubmark These properties make Rydberg atoms especially well suited to explore many-body quantum phenomena such as the recently experimentally realized symmetry-protected topological phases \cite{Leseleuc2019} and quantum dimer models \cite{Semeghini2021}.

Particularly interesting and still poorly understood many-body phases of spin systems are those where zero-point quantum fluctuations prevent magnetic order of any kind, which often requires 
frustration. 
The possibility of such liquid ground states 
were first pointed out by Philip Anderson in his seminal 1973 work on antiferromagnets \cite{Anderson1973,anderson1987}. Since then } there is an ongoing search for experimental realizations of such a quantum spin liquid (QSL) \cite{mila2000quantum,lee2008end,Savary2016,broholm2020quantum,Shen2016,Coldea2001,Liu2022,Gohlke2018}.
QSL may be gapped or gapless
and in the first case can be topological, where the topological order can either be protected by symmetries
 associated with short-range entanglement \cite{Lu-PRB-2012,Wen-PRB-2013}, or can be intrinsic, in which case the state
is long-range entangled \cite{chen2010local,Wen-NatScRev-2016}.  Despite a decades-long interest in quantum spin liquids, their clear identification in realistic materials remains a major challenge due to the scarcity of highly entangled states in real solid-state materials and the lack of simple experimental signatures of spin liquids \cite{Knolle2019,Wen2019,Norman2016}.
Thus it is natural to ask if QSLs can be realized in experimentally accessible model systems such as {\resubmark arrays of Rydberg atoms}. First signatures of a QSL have 
{ indeed }been found in such a model system in a recent breakthrough experiment with Rydberg atoms in an artificially assembled two-dimensional array of micro-traps \cite{Semeghini2021}. Here ground and Rydberg states of the atom form a spin-$1/2$ system.
The atoms were placed on the links of a Kagome lattice and driven under conditions of Rydberg blockade \cite{PhysRevLett.87.037901}, which effectively realizes a quantum dimer model for which a QSL ground state has been predicted  {\cite{Verresen2021, samajdar2021quantum}}. The dimer states
are formed by three nearby atoms out of which at most one can be excited due to
Rydberg blockade.

%%%%%%%%%%%%%%%%%%%%%%%%%%%%%%%%%%%%%%%%%%%%%%%%%%%%%%
\begin{figure}[htb]
	\centering
	\includegraphics[width=\columnwidth]{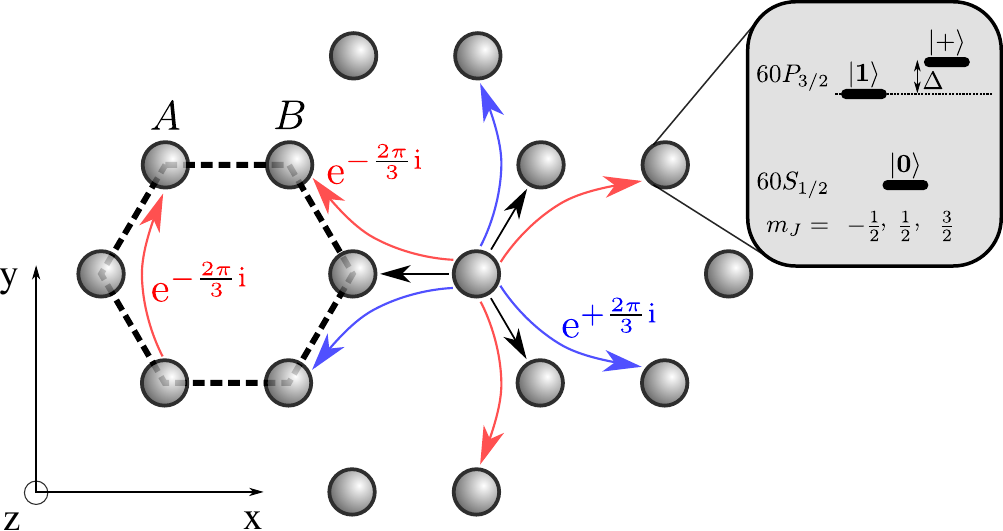}
	\caption{Honeycomb lattice with a two-site unit cell (A and B) of trapped atoms excited to two different Rydberg states $\ket{1}$ and $\ket{0}$, forming spin-$1/2$ systems. As indicated, spin-orbit coupling induced by an external magnetic field leads to nonlinear, complex {(chiral)} second-order hopping processes to the next-nearest neighbor (NNN) in addition to direct nearest neighbor (NN) hopping. The relevant level structure of a single atom is shown in the inset. The NNN hopping is facilitated by virtual transitions from $\vert 0\rangle$ to the off-resonant state $\vert +\rangle$ and can be controlled by varying its detuning $\Delta$. 
	}
	\label{fig:RydbergHaldane_allHopping}
\end{figure}
%%%%%%%%%%%%%%%%%%%%%%%%%%%%%%%%%%%%

Inspired by recent experimental work \cite{Lienhard2020} we here {\resubmark propose and analyze a lattice spin model, based on Rydberg atoms on a honeycomb lattice as shown in Fig.~\ref{fig:RydbergHaldane_allHopping},
where the competition between nearest and next-nearest $XY$ spin couplings leads to frustration.} 
Different from
\cite{Semeghini2021} the spin degree of freedom is formed by two Rydberg states of the atoms. The spin can 
hop from one lattice site to the next by dipolar exchange interactions.
As shown  in \cite{Lienhard2020}, in such systems spin-orbit coupling induced by an external magnetic field explicitly breaks time-reversal and chiral symmetry and leads to a density-dependent, second-order complex hopping of excitations \cite{Goerg2019} which competes with the direct hopping. The strength of the complex hopping can be modified by tuning the energy separation between the Rydberg states. Different from \cite{Semeghini2021} the Hamiltonian describing the system conserves the number of excitations, i.e. it has $U(1)$ symmetry, and we here consider half filling. 
We study the system using numerical simulations performing exact diagonalization (ED) on small lattices with {up to 30 lattice sites} with periodic boundary conditions using different cluster shapes.

If the effects of quantum fluctuations on the complex hopping are neglected, i.e. in mean-field approximation, there {\resubmark is a competition between the nearest-neighbor (NN) hopping, which tries to establish a condensate of the hard-core bosons, and the next-nearest-neighbor (NNN) hopping, driving the system into a 120$^\circ$ or spiral spin phase.}
A transition between the { two phases} occurs if the strength of the mean-field second-order hopping becomes comparable to the direct one. 

In the full model we identify two new phases. In the vicinity of the mean-field critical point an intermediate phase emerges, which is bare of any simple spin order, has a non-vanishing spin chirality,  and is characterized by a non-trivial many-body Chern number $C=1$. Thus this phase is a candidate for a topological QSL. {\resubmark Its precise nature is however not completely clear and needs further  investigations \cite{Marcello}. ED simulations for finite systems point to a gapped,
non-degenerate ground state. The many-body gap does however vary in magnitude  substantially when different twisted boundary conditions are applied.}
A gapped and non-degenerate 
ground state would indicate symmetry-protected topological (SPT) order. On the other hand the Chern number of $C=1$, which is robust to potential disorder, is odd and thus different from the even values found in other systems showing a symmetry-protected bosonic interger quantum Hall (BIQH) effect \cite{Lu-PRB-2012,Senthil-PRL-2013,Sterdyniak-PRL-2015,He2015,Liu-PRB-2019} and expected from general classification arguments \cite{Lu-PRB-2012}.
{\resubmark An odd value of the Chern number would on the other hand be consistent with a gapless QSL. In this case the collective gap should vanish in the thermodynamic limit and the ground state should become degenerate.}

When the second-order hoppings have the opposite sign and become strong another
disordered spin phase emerges, which is again chiral but gapless. We believe that this phase is a candidate for a gapless chiral spin liquid but the parameter regime is outside of what can be realized in Rydberg systems. 

Hard-core bosons on a honeycomb lattice with frustrated next nearest neighbor hopping have been studied in 
\cite{Varney2011,Varney2012} where {\resubmark the authors reported} evidence for a particular type of gapless spin liquids, called Bose metal, i.e. a QSL with a "Fermi-like" surface in momentum space. {\resubmark Subsequent DMRG simulations however showed weak density order \cite{zhu2013unexpected}, which demonstrates that an umambiguous identification of the nature of spin liquids using ED simulations is difficult.}
\\
\\
The outline of this publication is as follows: In Sec.\ref{sect:RydbergHam_HoneyLattice} we introduce the many-body Hamiltonian of Rydberg spin excitations in a two-dimensional honeycomb array of trapped atoms. An overview of the ground-state phases is given in Sec.\ref{sec:ground-state_phases} where we show that the presence of complex and density-dependent second-order hopping processes 
%of Rydberg excitations 
gives rise to two disorderd spin phases in addition to a trivial BEC phase and a spiral, or $\ang{120}$ phase present in a mean-field Hamiltonian. {\resubmark We comment on the possible nature of the QSL in Sec.\ref{sect:nature} and discuss the effects of longer-range interactions
in Sec.\ref{sect:long-range}.} 
Finally a summary and discussion of the results is given in Sec.\ref{sec:summary}.

%%%%%%%%%%%%%%%%%%%%%%%%%%%%%%%%%%%%%%%%%%%%%%%%%%%%%%
\section{Model for Rydberg Excitations on a Honeycomb Lattice}\label{sect:RydbergHam_HoneyLattice}
%%%%%%%%%%%%%%%%%%%%%%%%%%%%%%%%%%%%%%%%%%%%%%%%%%%%%%

%
We consider a honeycomb array of micro-traps filled with one atom each as shown in Fig.~\ref{fig:RydbergHaldane_allHopping}. 
Each site has three nearest neighbors (NN) of the opposite and six next-nearest neighbors (NNN) of the same sublattice A and B, respectively. As has been demonstrated in recent works \cite{Endres2016,Bernien2017,Barredo2016,Barredo2018}, a deterministic and defect-free preparation of such lattice structures with characteristic separations in the $\mu$m scale is possible and state of the art.
Each atom is excited into high lying Rydberg states, e.g.
within the 60$S_{1/2}$ and 60$P_{3/2}$ manifold of $^{87}$Rb as in \cite{Lienhard2020}. Application of an external magnetic field perpendicular to the plane leads to a level structure where only three  magnetic
sublevels are relevant, indicated in Fig.~\ref{fig:RydbergHaldane_allHopping}. Two of them, here labelled as $\vert 0\rangle$
and $\vert 1\rangle$, form an effective spin $1/2$ systems, and we are interested in the many-body dynamics of these
spins. Dipolar coupling between the Rydberg-excited atoms leads to a hopping of spin excitations ($XY$ coupling)
with an amplitude $J$ proportional to $1/r^3$, with $r$ being the atom separation.  
In the case of a transition between the $m_J=\pm \frac{1}{2}$ sublevels of $S_{1/2}$ and $P_{3/2}$, as mentioned above, $J=d^2/(8\pi\epsilon_0 r^3)$, with $d$ being the dipole matrix element between
states $\vert 0\rangle$ and $\vert 1\rangle$.
The third level, denoted by $\vert +\rangle$ is used to facilitate a second-order, off-resonant spin exchange process that is associated with a
geometry-dependent complex phase, see Fig.~\ref{fig:RydbergHaldane_allHopping}, and which depends on the spin state
of the intermediate atom. The microscopic physics of the system has been
studied both theoretically and experimentally in a minimal set-up \cite{weber2018topologically,Lienhard2020} and the relevant terms of the many-body Hamiltonian have been introduced and studied for a different lattice in \cite{Ohler2022}.
For a detailed derivation of the Hamiltonian we thus refer to these publications. Following along the same lines as in \cite{Ohler2022}, we write down the effective Hamiltonian for Rydberg excitations in the hard-core boson language (we use $\hbar=1$): 
\begin{align}\label{eqn:Rydberg_BHH_fullH}
\hat{H}=
&
-J\sum_{\langle i,j\rangle}\hat{b}_{j}^{\dagger}\hat{b}_{i}
-2gJ\sum_{\langle\langle i,j\rangle\rangle}\hat{b}_{j}^{\dagger}\hat{b}_{i}\mathrm{e}^{\pm\frac{2\pi\mathrm{i}}{3}}(1-\hat{n}_{ij})
\nonumber\\
&+2gJ\sum_{\langle i,j\rangle}\hat{n}_{i}\hat{n}_{j},
\end{align}
where $\mathrm{e}^{\pm\frac{2\pi\mathrm{i}}{3}}=-\frac{1}{2}\pm\frac{\sqrt{3}}{2}\mathrm{i}$ and $\hat{b}_{i}^{\dagger},\hat{b}_{i}$ create or destroy a hard-core boson on site $i$, respectively.
% The notations 
$\langle i,j\rangle$ and $\langle\langle i,j\rangle\rangle$ refer to NN and NNN, {\resubmark where both $i\to j$ and $j\to i$ are included in the sum.}
The sign of the complex phase as well as the intermediate site of the NNN hopping terms connecting sites $i$ and $j$ 
is indicated in Fig.~\ref{fig:RydbergHaldane_allHopping} by differently coloured and bent arrows, respectively. 
Thus the hopping between two nearest neighbors $i$ and $j$ of the same sublattice is controlled by a site of the opposite sublattice, located in between the two and with particle number $\hat{n}_{ij}$. 
Note that the complex phase picked up in a closed loop around a honeycomb plaquette corresponds to exactly one
flux quantum.
Furthermore, in the nearest-neighbor interaction term we have assumed conservation of particle number and dropped the constant energy-shift
\begin{align}
\sum_{\langle i,j\rangle}\hat{n}_{i}\left(1-\hat{n}_{j}\right)
\quad \rightarrow\quad
-\sum_{\langle i,j\rangle}\hat{n}_{i}\hat{n}_{j}.
\end{align}

All processes contained in the Hamiltonian are shown in Fig.~\ref{fig:RydbergHaldane_allHopping}. This includes 1) NN hopping with constant amplitude $J$ which depends on the atomic level structure and the
spatial separation between the atoms, 2) NNN hopping that is density-dependent, possesses a staggered complex phase and scales with an additional parameter $g$ and 3) NN interaction that also scales with $g$. Terms connecting sites further apart are smaller in magnitude and will be neglected in first approximation. 
We will discuss their influence at the end of the paper.
In eq.\eqref{eqn:Rydberg_BHH_fullH} the strength of the non-resonant processes $g$ is given by $g=27J/(2\Delta)$, where $\Delta$ denotes the detuning between two Rydberg states of the atoms. The factor $27$ stems from Clebsch-Gordan coefficients and factors in the microscopic Hamiltonian (see \cite{Ohler2022}). The additional factor $1/2$ in the definition of $g$ is introduced to be consistent with Ref.\cite{Ohler2022}, where the same atomic setup is studied on a zig-zag chain. Most importantly 
the magnitude and sign of $g$ can be controlled by the detuning of the internal state $\ket{+}$.
In order to be able to neglect population of the off-resonant state the detuning cannot be too small, i.e. $J/\vert \Delta\vert \ll 1$, but values of $\vert g\vert \sim 2$ are possible. 
In the present paper we consider half filling of hard-core bosons corresponding to a vanishing total magnetization.

Let us first discuss some general aspects of Hamiltonian \eqref{eqn:Rydberg_BHH_fullH}.
The presence of complex hopping amplitudes means that time-reversal symmetry is explicitly broken. The
microscopic origin of this is the magnetic field used to select the specific sublevels of the Rydberg atoms. 
Secondly, without the nonlinear term in the NNN hopping the model is symmetric under a combined time-reversal and
particle-hole transformation at half filling but this symmetry is broken by the term $\left(1-\hat{n}_{ij}\right)$.
The NN density-density interaction corresponding to a ferromagnetic $\left(gJ<0\right)$ or anti-ferromagnetic $\left(gJ>0\right)$ Ising term
would drive the system into a density-ordered state. The NNN hopping which is of the same strength however prevents the formation of a state with ferromagnetic or antiferromagnetic density order. Therefore, the possible phases are essentially governed by the competition of the NN and NNN hopping terms and the action of the nonlinear term in the NNN hopping amplitude.

%%%%%%%%%%%%%%%%%%%%%%%%%%%%%%%%%%%%%%%%%%%%%%%%%%%%%%%%%%%%%%%%%%%%%%%%%%%
\section{Ground-state phases and Effects of Nonlinear Hopping}\label{sec:ground-state_phases}
%%%%%%%%%%%%%%%%%%%%%%%%%%%%%%%%%%%%%%%%%%%%%%%%%%%%%%%%%%%%%%%%%%%%%%%%%%%

\begin{figure}[htb]
	\centering
	\includegraphics[width=0.5\textwidth]{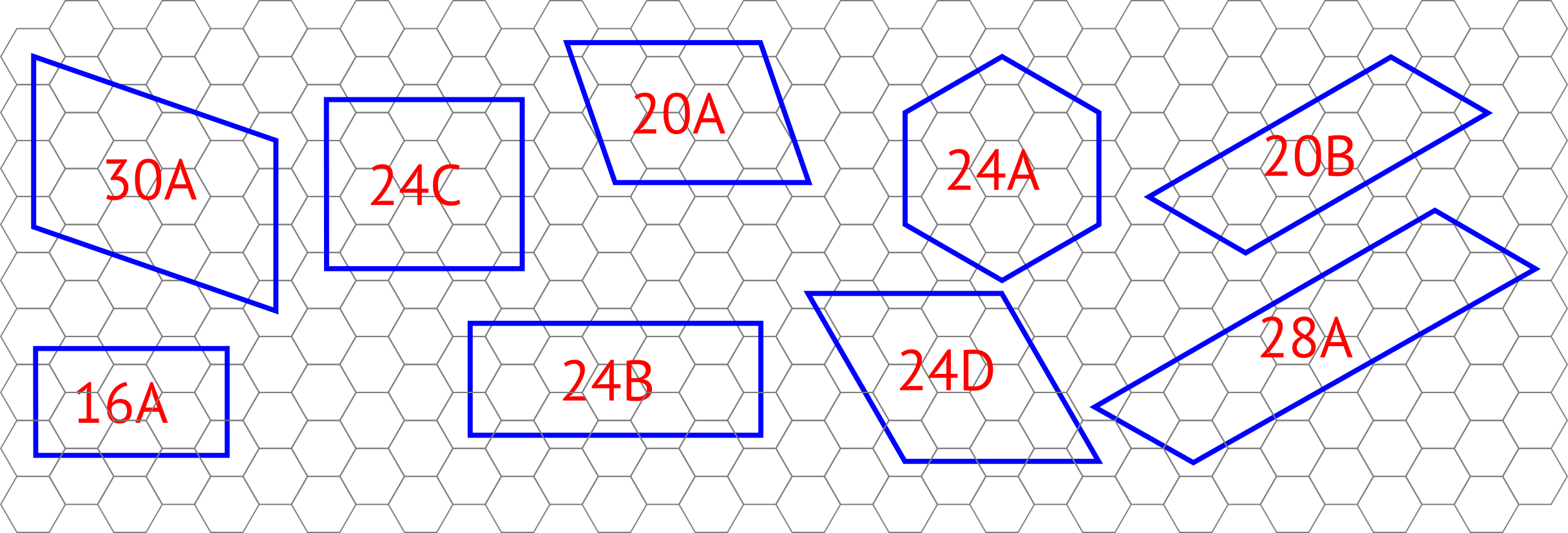}
	\caption{Collections of cluster shapes and sizes used for numerical calculations. The nomenclature is consistent with \cite{Varney2012}.}
	\label{fig:TwoD_ShapeCluster}
\end{figure}

In order to investigate the different ground-state phases of the model \eqref{eqn:Rydberg_BHH_fullH} we use exact diagonalization (ED) on finite lattices using periodic (or twisted) boundary conditions. In order to reduce boundary effects, we perform calculations on hexagonal clusters of varying shapes and sizes. The clusters that we use are shown in Fig.~\ref{fig:TwoD_ShapeCluster}. Using the Lanczos algorithm \cite{Virtanen2020short}, we gain access to the ground state wave function.
 %
%%%%%%%%%%%%%%%%%%%%%%%%%%%%%%%%%%%%%%%%%%%%%%%%%%%%%%%%%%%
%%%%%%%%%%%%%%%%%%%%%%%%%%%%%%%%%%%%%%%%%%%%%%%%%%%%%%%%%%%%
%

To obtain a general overview of the phase diagram as a function of the parameter $g$ we consider the change of the ground-state wavefunction $\ket{\Phi_0}$ upon infinitesimal changes of $g$. We can quantify this by the dimensionless, intensive fidelity metric 
\begin{equation}
    f(g)=\frac{2}{\systemsize} \frac{1-\abs{\bra{\Phi_0(g)}\ket{\Phi_0(g+\delta g)}}}{\left(\delta g\right)^2},\qquad \delta g\to 0.
    \label{eqn:fidelity_definition}
\end{equation}
Here, $\systemsize$ represents the number of sites in the system. This quantity has been shown to be a useful indicator of quantum phase transitions \cite{Zanardi2006} and has since been used in numerous condensed-matter applications \cite{CamposVenuti2008,Yang2007,Varney2011,Varney2012}. By computing the overlap of the ground-state wavefunction with itself under small changes of $g$, we are able to detect the regions in parameter space where the system's ground-state changes rapidly, indicating a possible quantum phase transition (QPT). In a finite system this quantity will always be finite and 
not show a Dirac-$\delta$-like behaviour which we expect at the critical point of a QPT in an infinite system. Therefore we can use it only as a rough guide to separate different parameter regimes.

As we will show, the density-dependent hopping in \eqref{eqn:Rydberg_BHH_fullH} has a profound impact on the behaviour of the system. 

%%%%%%%%%%%%%%%%%%%%%%%%%%%%%%%%%%%%%%%%%%%%%%%%%%%%%%
\subsection{{\resubmark Mean-field Dynamics: Competition between NN and NNN Hopping}}\label{subsect:mean_field_approx}
%%%%%%%%%%%%%%%%%%%%%%%%%%%%%%%%%%%%%%%%%%%%%%%%%%%%%%

{\resubmark To gain some insight about the main competing terms in the Hamiltonian we first study a 
mean-field approximation. To this end we drop the nearest-neighbor interaction term and} replace the density-dependence of the NNN hopping term with a constant expectation value
\begin{align}
    \left(1-\hat{n}_{ij}\right)
    \to
    \left(1-\bar{n}\right).
\end{align}
Here $\bar{n}$ denotes the average density of the lattice. Since we consider half filling we set  $\bar{n}=0.5$. The modified approximate Hamiltonian then reads
{\resubmark
\begin{align}
\hat{H}_{MF}=
&
-J\sum_{\langle i,j\rangle}\hat{b}_{j}^{\dagger}\hat{b}_{i}
-gJ\sum_{\langle\langle i,j\rangle\rangle}\hat{b}_{j}^{\dagger}\hat{b}_{i}\mathrm{e}^{\pm \frac{2\pi}{3}\mathrm{i}}
\label{eqn:Rydberg_BHH_MFinHopping}
\end{align}
}
This mean-field Hamiltonian is that of the Bose-Hubbard Haldane model. The corresponding single-particle band-structure is that of Haldane's generalization of graphene in the topologically non-trivial 
regime \cite{Haldane1988}. However, as shown in \cite{Varney2010}, the hard-core boson character of the Rydberg excitations makes
the many-body ground state topologically trivial. Note that while in one dimension hard-core bosons and fermions can behave identically under certain circumstances, this is decidedly not the case in two dimensions and the generalization
of the Wigner-Jordan transformation to two dimensions requires to introduce effective gauge fields \cite{Fradkin1989}. 
%
%%%%%%%%%%%%%%%%%%%%%%%%%%%%%%%%%%%%%%%%%%%%%%%%%%%%%%%%%%%%%
\begin{figure}[htb]
	\centering
    \includegraphics[width=0.95\columnwidth]{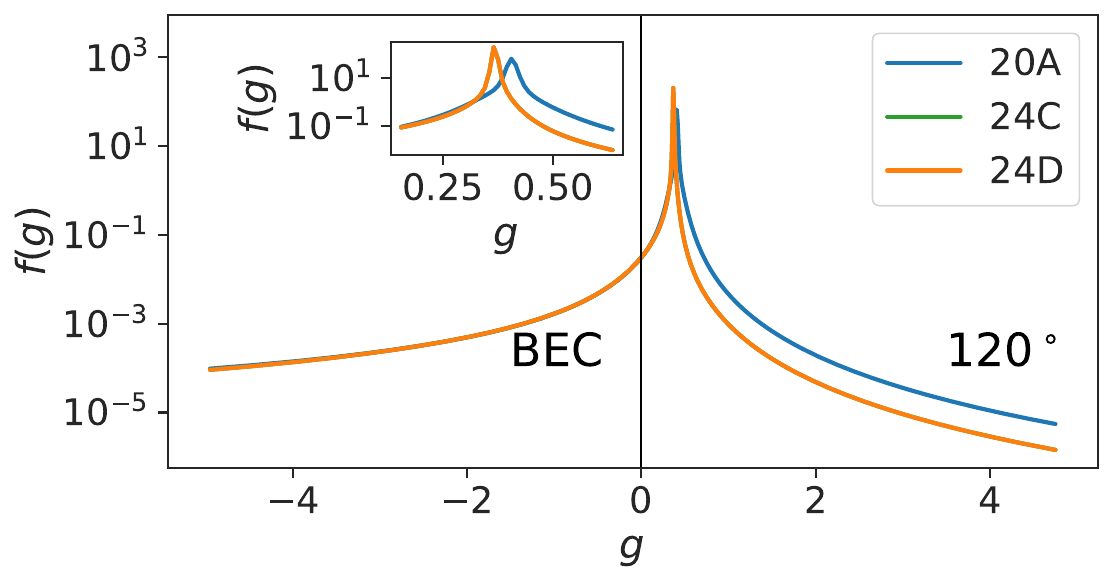}
	\caption{Ground-state fidelity metric $f$ for the mean-field system given in \eqref{eqn:Rydberg_BHH_MFinHopping} as a function of the parameter $g$. The peak of $f$ agrees well for the different system shapes and indicates a phase transition. For $g\lesssim {\resubmark 0.4}$ we find a superfluid state, while for $g\gtrsim {\resubmark 0.4}$ the system shows $\ang{120}$-order. {\resubmark The inset shows the region around the phase transition in more detail. Note that the curves for shapes 24C and 24D are virtually indistinguishable.}}
	\label{fig:GS_fidelity_shapes_MF}
\end{figure}
%%%%%%%%%%%%%%%%%%%%%%%%%%%%%%%%%%%%%%%%%%%%%%%%%%%%%%%%%%%%%
%
In Fig.~\ref{fig:GS_fidelity_shapes_MF} we have plotted the fidelity metric of the ground-state as a function of the interaction strength $g$, using the mean-field Hamiltonian \eqref{eqn:Rydberg_BHH_MFinHopping} with periodic boundary conditions (PBC) on a torus. From the fidelity we see two regimes separated by a single peak. The left region is continuously connected to the trivial limit $g=0$, where the system is in a BEC state. In order to understand the phase in the right region it is sufficient to consider the case $g\gg 1$ {\resubmark where the NN hopping term is irrelevant}
{\resubmark
\begin{align}
\hat{H}_{g\to\infty}=
&
-gJ\sum_{\langle\langle i,j\rangle\rangle}\hat{b}_{j}^{\dagger}\hat{b}_{i}\mathrm{e}^{\pm\frac{2\pi}{3}\mathrm{i}}
\label{eqn:Rydberg_BHH_120trial}
\end{align}
}
Now, the two triangular sublattices $(A)$ and $(B)$ of the hexagonal lattice are disconnected  and the internal dynamics in each of the
sublattices is determined by the NNN hopping. Furthermore, we can rewrite the Hamiltonian in terms of spin-$1/2$ matrices
\begin{align}\label{eqn:S_ops_from_boson_ops}
\hat{b}^{\dagger}\rightarrow\hat{S}^{+}=\hat{S}^{x}+\mathrm{i}\hat{S}^{y}
&&
\hat{b}\rightarrow\hat{S}^{-}=\hat{S}^{x}-\mathrm{i}\hat{S}^{y}
\end{align}
such that the NNN hopping term reads
\begin{align}
\hat{H}_{{g\to\infty}}=
-\frac{gJ}{2}\sum_{\triangle}
\big(
    \vec{S}_{\triangle,1}^{T}
    &
    D\vec{S}_{\triangle,2}
    +
    \vec{S}_{\triangle,2}^{T}D\vec{S}_{\triangle,3}\notag
    \\
    &+
    \vec{S}_{\triangle,3}^{T}D\vec{S}_{\triangle,1}
\big).
\end{align}
Here, the  vector operators $\vec{S}_{\triangle,i}$ are projections of the spins to the $xy$ plane, the index $\triangle$ runs over all triangles of both triangular sublattices and the index $1,2,3$ iterates through a single triangle as indicated in the insert of Fig.~\ref{fig:GS_fidelity_shapes_MF}. The matrix $D=D\left(2\pi/3\right)$ is the rotational matrix around the $z$-axis with rotation angle of $2\pi/3$. With $D^{3}=1$ we can write $\hat{H}_{{g\to\infty}}$ in its final form
\begin{align}
\hat{H}_{{g\to\infty}}=
&
-\frac{gJ}{2}\sum_{\triangle}
\left(\vec{S}_{\triangle}+D\vec{S}_{\triangle,2}
    +D^{2}\vec{S}_{\triangle,3}
\right)^{2}
\\
&\hspace{1cm}+\mathrm{const.}\notag
\end{align}
For $g>0$ we see that in the ground state the rotated spins $\vec{S}_{\triangle,1}$, $D\vec{S}_{\triangle,2}$ and $D^{2}\vec{S}_{\triangle,3}$ need to be parallel, i.e. the spin vectors themselves have to be at an angle of $\ang{120}$ in the $xy$-plane. We can confirm this calculation by considering the in-plane spin correlations, which we define as
\begin{align}\label{eqn:inplanecorr}
    \inplanecorr{\theta}=4\expval{\hat{S}_{i}^{(0)}\hat{S}_{j}^{(\theta)}},
\end{align}
where
\begin{align}
    \hat{S}_{j}^{(\theta)}=
    \cos(\theta)\hat{S}^{x}_{j}+\sin(\theta)\hat{S}^{y}_{j}.
\end{align}
This correlation function detects if both spin vectors are separated by an angle $\theta$ in the equatorial plane. The factor of four is introduced to normalize $\inplanecorr{\theta}$ to unity in case of a perfectly correlated state. 
%
%%%%%%%%%%%%%%%%%%%%%%%%%%%%%%%%%%%%%%%%%%%%%%%%%%%%%%%%%%%%%%%%%%
\begin{figure}[htb]
	\centering
	\includegraphics[width=\columnwidth]{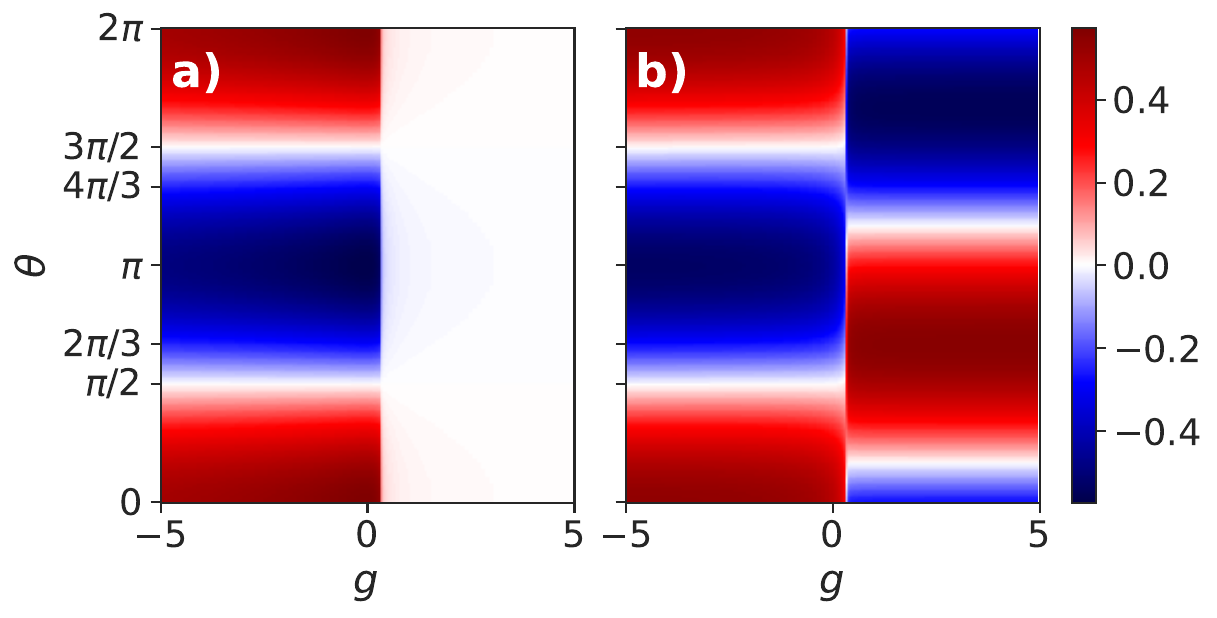}
	\caption{In-plane spin-spin correlation \eqref{eqn:inplanecorr}, in mean-field approximation \eqref{eqn:Rydberg_BHH_MFinHopping}. The left figure shows $\inplanecorr{\theta}$ for $i$ and $j$ being NN (different sublattice), the right figure depicts NNN (same sublattice) correlation. All calculations were performed with periodic boundary conditions on the shape 24C.  The $\ang{120}$-regime as well as the parallel alignment in the BEC phase can be clearly seen. Furthermore one recognizes that {\resubmark NNs are completely uncorrelated in the $\ang{120}$-phase}.}
	\label{fig:SpinOrientation_Correlation_MF}
\end{figure}
%%%%%%%%%%%%%%%%%%%%%%%%%%%%%%%%%%%%%%%%%%%%%%%%%%%%%%%%%%%%%%%%%%%

In Fig.~\ref{fig:SpinOrientation_Correlation_MF} we show the results for the in-plane spin correlations in the mean-field limit, which show a maximum at $\theta=\ang{120}$ for large $g$. The case of $g<0$ can be understood similarly: Here, the sum over the rotated spin vectors has to vanish, which requires the spins to align in a parallel way in the $xy$ plane. The $\theta=\ang{120}$ phase has a remaining $SO(2)$ symmetry. {\resubmark The lack of any correlation between NNs in the $\ang{120}$-phase for large $g$ is a trivial result of the disconnected sub-lattices in this limit}. The small deviation of $C(\ang{120})$ from the maximum value of $1/2$ (taking into account the $SO(2)$ symmetry) can be attributed to finite size effects.

%%%%%%%%%%%%%%%%%%%%%%%%%%%%%%%%%%%%%%%%%%%%%%%%%%%%%%%%%%%%%%%%%%%
\subsection{Full Hamiltonian}

\begin{figure}[htb]
	\centering
    \includegraphics[width=0.95\columnwidth]{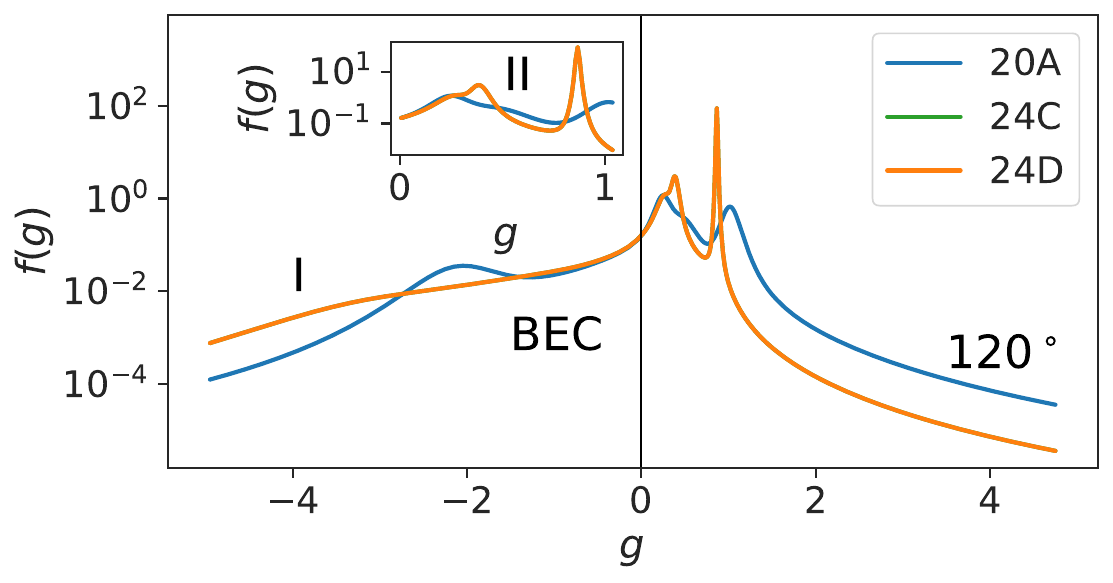}
	\caption{Ground-state fidelity metric $f$ as a function of the parameter $g$ for the full Hamiltonian \eqref{eqn:Rydberg_BHH_fullH}. The peaks of $f$ agree well for the different shapes and indicate potential phase transitions. The inset shows regime $\mathrm{II}$ in detail. The BEC and $\ang{120}$-order regimes agree with the mean-field model, but two new regimes appear, which we label $\mathrm{I}$ and $\mathrm{II}$.}
	\label{fig:GS_fidelity_shapes_full}
\end{figure}

%%%%%%%%%%%%%%%%%%%%%%%%%%%%%%%%%%%%%%%%%%%%%%%%%%%%%%%%%%%%%%%%%%%
After having discussed the mean-field Hamiltonian, we now turn to the microscopically motivated full Hamiltonian \eqref{eqn:Rydberg_BHH_fullH}, which includes a density-dependent, complex NNN hopping term and {NN density-density interaction}. In this case, the ground-state fidelity \eqref{eqn:fidelity_definition} is modified substantially, as shown in Fig.~\ref{fig:GS_fidelity_shapes_full}. Where for the mean-field model we had seen only one phase transition at $g$ close to $0.4$, we now see two sharp peaks of the fidelity in the vicinity of the mean-field critical point. Thus a new phase emerges in between the BEC ($g\to 0$) and the $\ang{120}$-order ($g\gg 0.4$), which we refer to as regime $\mathrm{II}$. Additionally, for $g\approx -5$ we see a behaviour that is qualitatively different from the mean-field case. Due to the lack of a clear peak in that region it is unclear whether this indicates another true phase transition or a crossover. This area of the parameter space we denote by $\mathrm{I}$.
We note, however, that in this region the condition $J/\vert \Delta\vert \ll 1$ resulting from microscopic physics of Rydberg interactions and required to neglect
the population in level $\vert +\rangle$ is no longer fulfilled.
%
%%%%%%%%%%%%%%%%%%%%%%%%%%%%%%%%%%%%%%%%%%%%%%%%%%%%%%%%%%%%%
%%%%%%%%%%%%%%%%%%%%%%%%%%%%%%%%%%%%%%%%%%%%%%%%%%%%%%%%%%%%%
%

In the following we will characterize the new phases by different observables.

%%%%%%%%%%%%%%%%%%%%%%%%%%%%%%%%%%%%%%%%%%%%%%%%%%
\subsubsection{In-Plane Spin Orientation}
%%%%%%%%%%%%%%%%%%%%%%%%%%%%%%%%%%%%%%%%%%%%%%%%%%%

\begin{figure}[htb]
	\centering
	\includegraphics[width=\columnwidth]{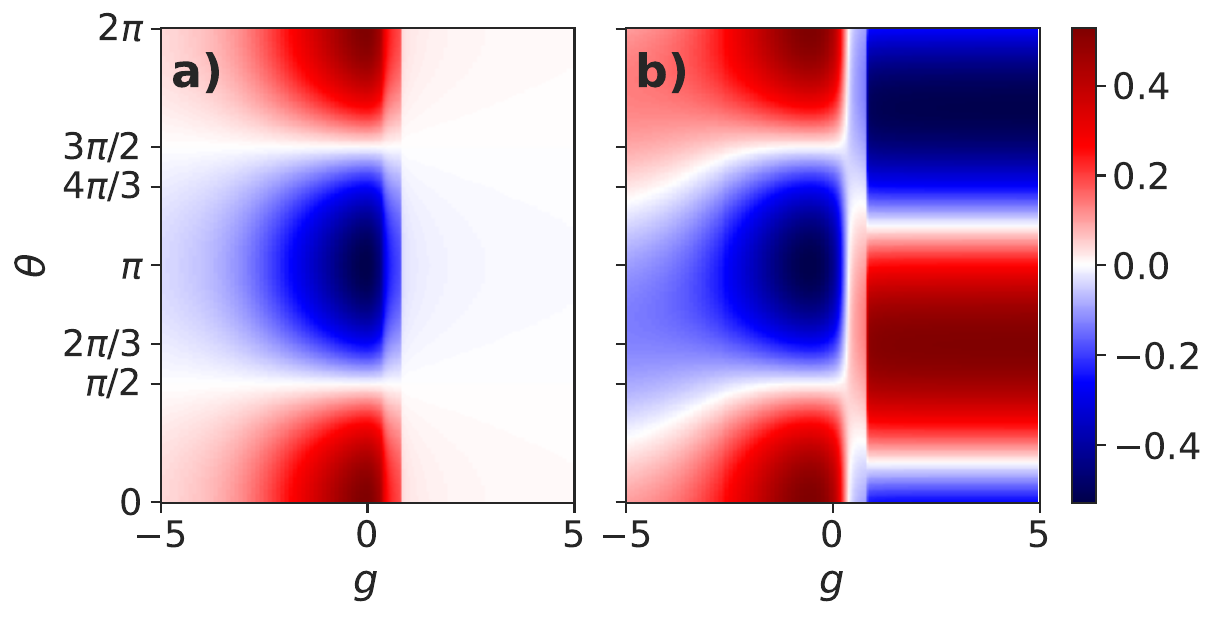}
	\caption{In-plane spin-spin correlation \eqref{eqn:inplanecorr} as in Fig.~\ref{fig:SpinOrientation_Correlation_MF} but for the full Hamiltonian \eqref{eqn:Rydberg_BHH_fullH}. One recognizes a very similar behavior for the BEC and $\ang{120}$ phases as in the mean field case. In phases I and II the in-plane spin correlations are suppressed.}
	\label{fig:SpinOrientation_Correlation_Full}
\end{figure}

In Fig.~\ref{fig:SpinOrientation_Correlation_Full} we show the results of the in-plane spin correlations
for the full Hamiltonian \eqref{eqn:Rydberg_BHH_fullH}. In the BEC as well as in the $\ang{120}$ phase the correlations are almost identical to the mean field case, supporting our interpretation of these phases. In the narrow intermediate phase II the in-plane spin correlations are suppressed and they are also reduced when entering and getting deeper into phase I. As opposed to the transition points to phase II the changes in the correlations at the transition into phase I are not sharp. 

%%%%%%%%%%%%%%%%%%%%%%%%%%%%%%%%%%%%%%%%%%%%%%%%%%%%%%%%%%%%%%%%%%
%%%%%%%%%%%%%%%%%%%%%%%%%%%%%%%%%%%%%%%%%%%%%%%%%%%%%%%%%%%%%%%%%%%

%%%%%%%%%%%%%%%%%%%%%%%%%%%%%%%%%%%%%%%
\subsubsection{Spin chirality}
%%%%%%%%%%%%%%%%%%%%%%%%%%%%%%%%%%%%%%%

\begin{figure}[htb]
	\centering
		\includegraphics[width=0.25\textwidth]{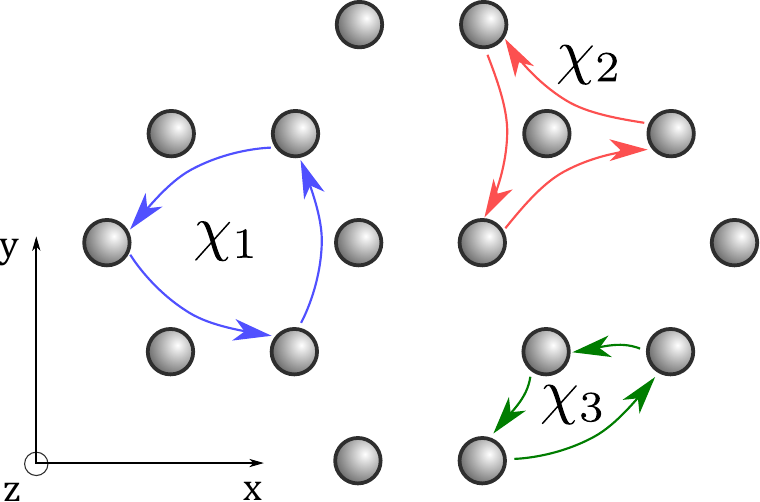}
		\
		
		\
		
		\includegraphics[width=0.48\textwidth]{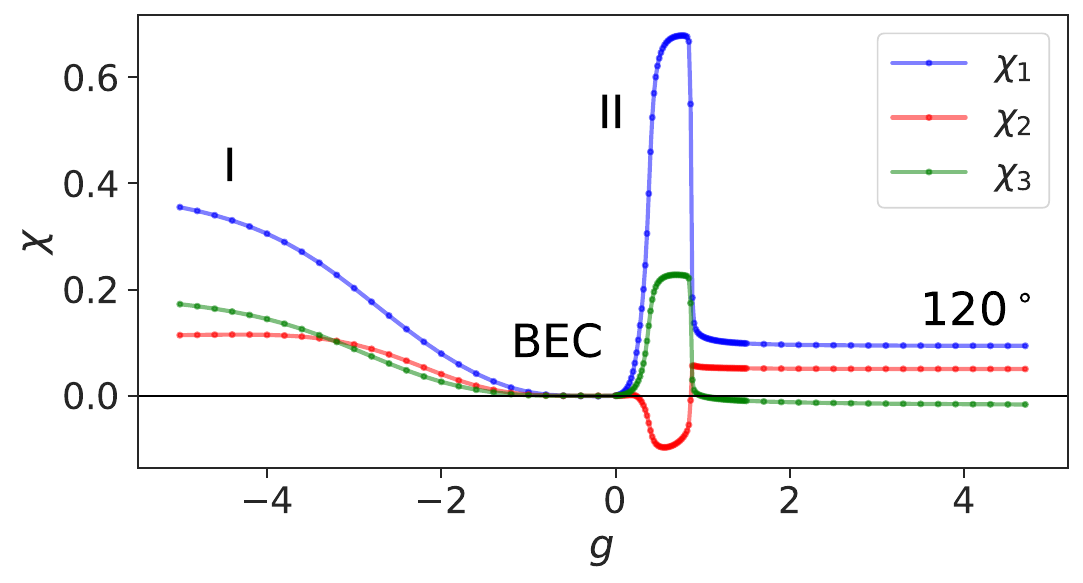}
	\caption{Visualization of the spin chirality measured for the full Hamiltonian \eqref{eqn:Rydberg_BHH_fullH}. Top: We denote three different types of triangles by $\chi_{i}$, $i\in\{1,2,3\}$. Bottom: Calculation of the spin chirality for the triangles displayed above. We observe that in regime $\mathrm{I}$ all three chiralities behave similarly, while in regime $\mathrm{II}$ the values differ in sign.}
	\label{fig:Spin_chirality_both}
\end{figure}

The Peierls phases in the NNN hopping terms explicitly break time-reversal symmetry. Due to the absence of a mass term shifting the energy of the A sub-lattice relative to that of the B sub-lattice, the mean-field Hamiltonian, eq.\eqref{eqn:Rydberg_BHH_MFinHopping}, preserves chiral symmetry, which amounts to a combination of time-reversal and particle-hole
transformation. The nonlinear term in the complex NNN hopping amplitude of the full model,
eq.\eqref{eqn:Rydberg_BHH_fullH},  however breaks the chiral symmetry.
Thus we expect that the disordered phases I and II are characterized by a significant spin chirality. The latter is defined as \cite{Hickey2016}
\begin{align}
\chi=\expval{\hat{\vec{\sigma}}_{i}\cdot\left(\hat{\vec{\sigma}}_{j}\cross\hat{\vec{\sigma}}_{k}\right)},
\label{eqn:SpinChirality_general}
\end{align}
where $\hat{\vec{\sigma}}$ is the 3-component vector of Pauli operators in the spin-$1/2$ representation of the hard-core boson model. The indices $\{i,j,k\}$ in \eqref{eqn:SpinChirality_general} are labelled in counter-clockwise order around the elementary triangles of the honeycomb lattice as displayed in Fig.~\ref{fig:Spin_chirality_both}. Chiral symmetry would enforce $\chi=0$. 
In Fig.~\ref{fig:Spin_chirality_both} we plot the spin chirality on the three types of triangles as a function of the interaction strength $g$. We observe that in both disordered regimes their values are much larger than in the BEC and $\ang{120}$-order phases. In the phase labelled as I in Fig.~\ref{fig:GS_fidelity_shapes_full} all three $\chi_{i}$ behave similarly, whereas in regime $\mathrm{II}$ their values differ in sign, indicating that the mediated interactions between the sublattices play an important role in the physics of regime $\mathrm{II}$.

%
%%%%%%%%%%%%%%%%%%%%%%%%%%%%%%%%%%%%%%%%%%
%%%%%%%%%%%%%%%%%%%%%%%%%%%

%%%%%%%%%%%%%%%%%%%%%%%%%%%%%%%%%%%%%%%%%%
\subsubsection{Spin Order}
%%%%%%%%%%%%%%%%%%%%%%%%%%%%%%%%%%%%%%%%%%

{\paragraph{Spin structure factor --} In order to investigate the presence or absence of spin order in phases I and II we now } consider the spin structure factor, defined as
\begin{align}\label{eqn:SpinOrder_equation}
S\left(\vec{k}\right)=
\frac{1}{\systemsize}\sum_{i,j=1}^{N}\mathrm{e}^{-\mathrm{i}\vec{k}\cdot\left(\vec{r}_{i}-\vec{r}_{j}\right)}
\expval{\hat{\vec{S}}_{i}\cdot\hat{\vec{S}}_{j}},
\end{align}
where $\hat{\vec{S}}_{j}=\left(\hat{S}^{x}_{j},\hat{S}^{y}_{j},\hat{S}^{z}_{j}\right)^{T}$ is the full 3D spin vector and 
$\systemsize$ is the system size. 
{\resubmark In order to minimize the impact of finite-size effects it is useful to impose a more general type of boundary conditions.
Specifically we consider twisted boundary conditions,
where a particle picks up a phase upon hopping over the boundary. 
Therefore we average $S(\vec{k})$ over all low-energy configurations with twisted boundary conditions (see subsequent paragraph \ref{paragraph:RTBC}).}
%
%%%%%%%%%%%%%%%%%%%%%%%%%%%%%%%%%%%%%%%%%%%%%%%%%%%%%%%%%%%%%
\begin{figure}[htb]
	\centering
	\includegraphics[width=\columnwidth]{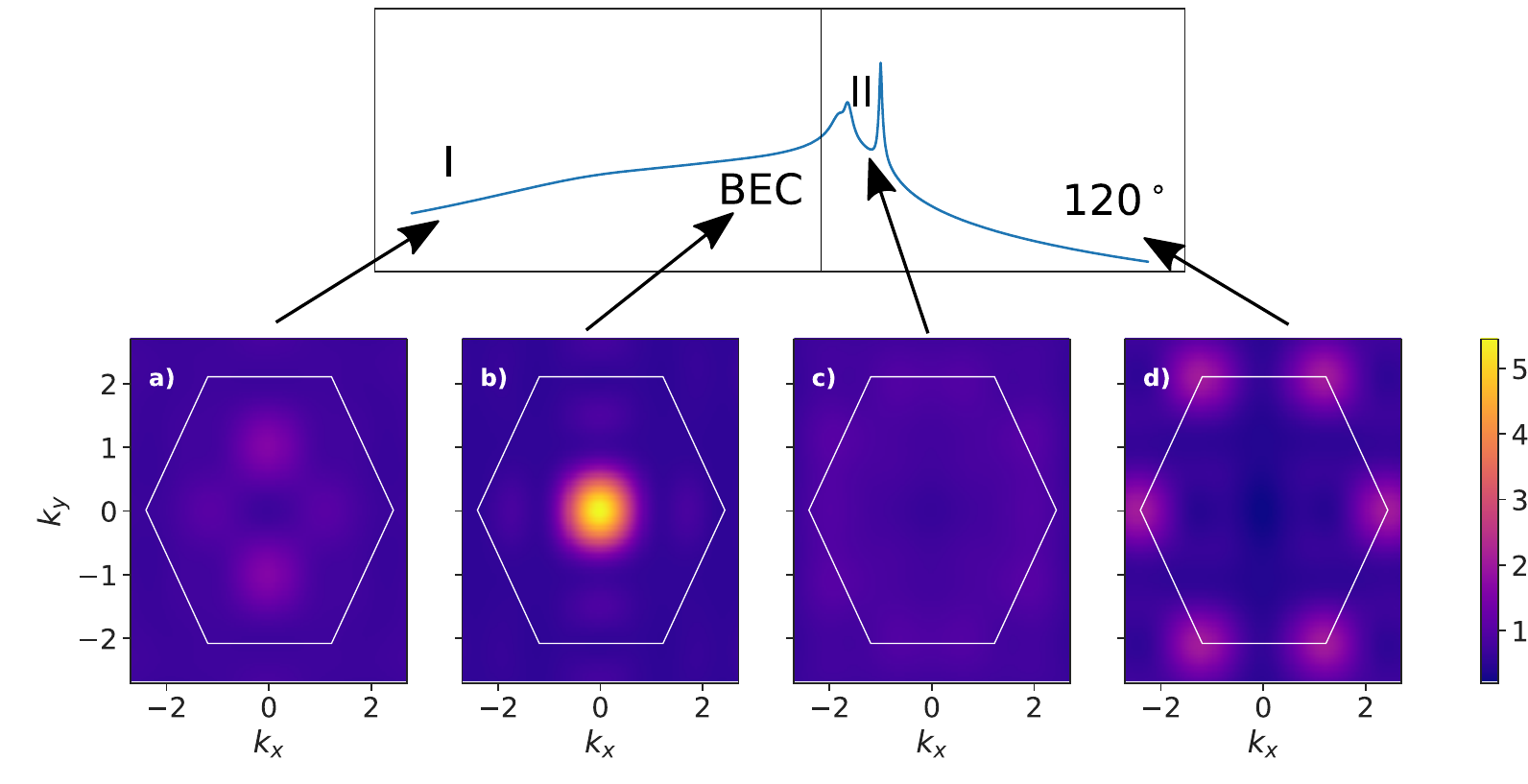}
	\caption{Averaged spin structure factor over all low-energy configurations for {(a) $g=-5.0$, (b) $g=0$, (c) {$g=0.73$} and (d) $g=4.0$}, respectively. For {$g=4.0$}, we see the well-known result for a $\ang{120}$-order, which is explained in the text. For {$g=-5.0$} and {$g=0.73$} we see no prominent features after the averaging.}
	\label{fig:SSSF_2d_plot}
\end{figure}
%%%%%%%%%%%%%%%%%%%%%%%%%%%%%%%%%%%%%%%%%%%%%%%%%%%%%%%%%%%%%
%

In Fig.~\ref{fig:SSSF_2d_plot} we show the results for the spin order, eq.~\eqref{eqn:SpinOrder_equation} for 
$L=24$ (shape 24C). In the BEC-regime ($g=0$) the plot shows a single prominent peak around zero momentum, consistent with parallel spins in the $xy$-plane as discussed before.
The $\ang{120}$-order occurs for $g>1$ such that NNN hopping terms and the density-density repulsion are dominant, whereas NN hopping is weak. Consequently, we can imagine the honeycomb lattice as two separate triangular sublattices. In this limit, the nearest neighbor spin-orientation in the $xy$-plane is completely uncorrelated, while NNNs align themselves at an angle of $\frac{2\pi}{3}=\ang{120}$. This is demonstrated in the hexagonal peaks of Fig.~\ref{fig:SSSF_2d_plot}d), which are a well-known indicator of $\ang{120}$-order \cite{CastellsGraells2019} (also referred to as spiral order). Additionally, as we have seen already in
Fig.~\ref{fig:SpinOrientation_Correlation_Full}, we find that the in-plane spin correlation \eqref{eqn:inplanecorr} vanishes for spins of different sublattices and prefers the angle $2\pi/3$ for the same sublattice. 

\begin{figure}
	\centering
		\includegraphics[width=0.8\columnwidth]{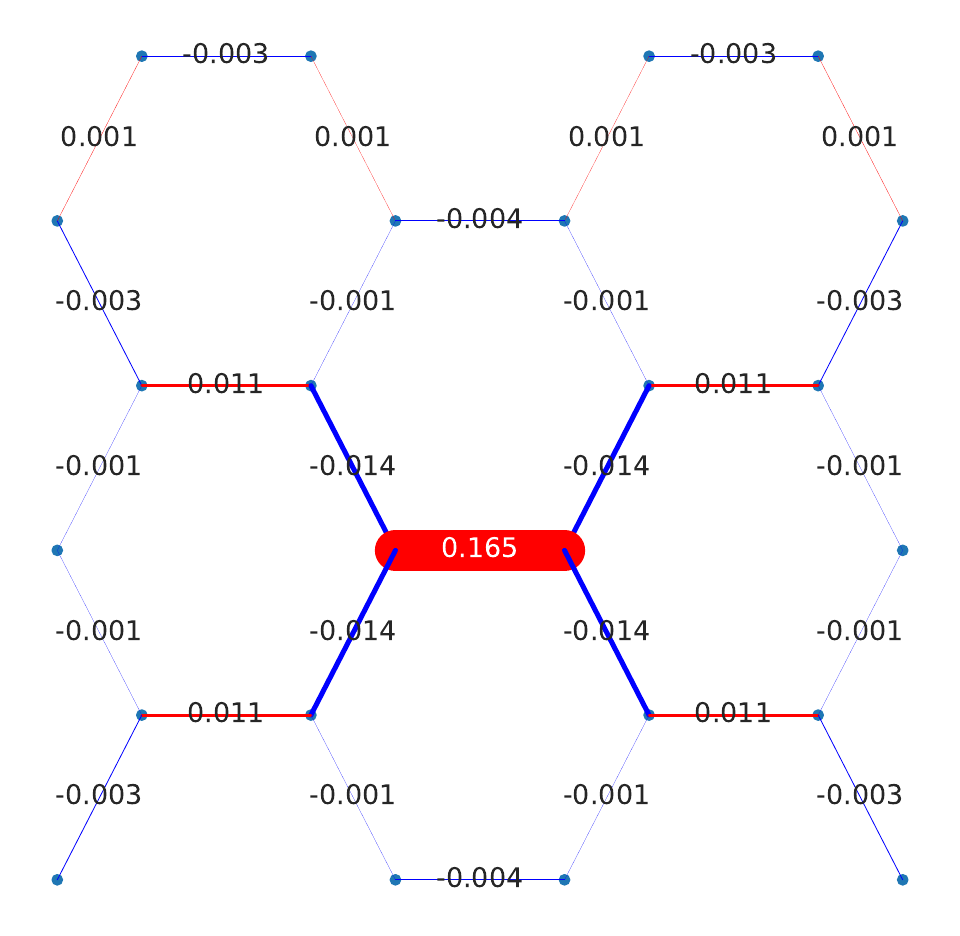}
	\caption{{\resubmark Dimerization $D_{ij, kl}$ as defined in \eqref{eqn:d_ij_kl} for $g=0.73$ (phase II). The linewidth and color encode the magnitude and sign, respectively. In this figure, we keep the reference bond $i,j$ constant and vary $k,l$. The reference bond color box is shown with rounded corners.}}
	\label{fig:dimerization_fullH}
\end{figure}

%\subsubsection{Dimerization}
{\resubmark \paragraph{Dimerization -- }
In order to test for the presence of more involved orderings we investigate if a regular pattern of dimerized spins occurs in the system. To this end we calculate the dimer-dimer correlation function introduced in \cite{meng2010quantum}
\begin{align}
    D_{ij, kl}=\expval{
    \left(
    \hat{\vec{S}}_{i}\cdot\hat{\vec{S}}_{j} - \frac{1}{4}
    \right)
    \left(
    \hat{\vec{S}}_{k}\cdot\hat{\vec{S}}_{l} - \frac{1}{4}
    \right)\nonumber
    }
    \\
    -
    \expval{\hat{\vec{S}}_{i}\cdot\hat{\vec{S}}_{j} - \frac{1}{4}}
    \expval{\hat{\vec{S}}_{k}\cdot\hat{\vec{S}}_{l} - \frac{1}{4}},\label{eqn:d_ij_kl}
\end{align}
where $i,j$ and $k,l$ are each NN. The results of the calculation are shown in Fig. \ref{fig:dimerization_fullH}, where we observe only short-ranged correlations.}
%%%%%%%%%%%%%%%%%%%%%%%%%%%%%%%%%%%
%%%%%%%%%%%%%%%%%%%%%%%%%%%

%\subsubsection{RTBC}
\paragraph{Randomly twisted boundary conditions -- }
\label{paragraph:RTBC} {\resubmark A rather general method to distinguish ordered from disordered phases in a quantum system is to study the reaction of the ground state to changes in boundary conditions. Therefore} we investigate the low-energy states of the model using randomly twisted boundary conditions (RTBC), following ideas introduced in \cite{Thesberg2014}. Specifically we study how changes by randomly twisting the boundary conditions affect the ground state. The twisting is performed by adding a complex phase to the hopping terms that cross a boundary in horizontal or vertical direction. A hard-core boson crossing the boundary in $x$-direction then acquires a phase $\theta_x$, for a vertical hop across the boundary it picks up a phase $\theta_y$, respectively, and the sum or difference $\theta_x\pm\theta_y$ in case of a diagonal crossing. 

For each realization, the phases $(\theta_x,\theta_y)$ are drawn at random from the uniformly distributed interval $\interval[open right]{0}{2\pi}$. For further reference we will use $P$ to denote the set of realizations, where
\begin{align}
\{\theta_{x}^{(p)},\theta_{y}^{(p)}\}
\in P,
\hspace{0.5cm}\forall 0\leq p\leq M,
\end{align}
and $M$ represents the number of realizations. Depending on the particular boundary condition in one realization, the ground-state energy $E\bigl(\theta_{x}^{(p)},\theta_{y}^{(p)}\bigr)$ and the ground-state vector $\ket{\Psi\bigl(\theta_{x}^{(p)},\theta_{y}^{(p)}\bigr)}$ will vary, some realizations resulting in higher or lower ground-state energies. Therefore, we define the \textit{optimal twist} $\left(\theta_{x}^{gs},\theta_{y}^{gs}\right)$ to be that realization which results in the minimal ground-state energy 
\begin{align}
E\left(\theta_{x}^{gs},\theta_{y}^{gs}\right)
\equiv
\min_{p\in P}
\bigg\{
E\left(\theta_{x}^{(p)},\theta_{y}^{(p)}\right)
\bigg\}.
\end{align}
Accordingly, we will refer to $\ket{\Psi\left(\theta_{x}^{gs},\theta_{y}^{gs}\right)}$ as the \textit{optimal ground-state vector} and to $E(\theta_{x}^{gs},\theta_{y}^{gs})$ as the \textit{optimal ground-state energy}. Then, we can normalize all energies $E(\theta_{x}^{(p)},\theta_{y}^{(p)})$ with respect to the optimal ground-state energy and compute the relative difference
\begin{align}
\epsilon_{p} = 
\frac{E\left(\theta_{x}^{(p)},\theta_{y}^{(p)}\right)-E\left(\theta_{x}^{gs},\theta_{y}^{gs}\right)}
{\abs{E\left(\theta_{x}^{gs},\theta_{y}^{gs}\right)}}>0,
\end{align}
as well as the overlap of each ground-state with the optimal ground-state
\begin{align}
O_{p}=\abs{
	\bra{\Psi\left(\theta_{x}^{(p)},\theta_{y}^{(p)}\right)}
	\ket{\Psi\left(\theta_{x}^{gs},\theta_{y}^{gs}\right)}}.
\end{align}
%
%%%%%%%%%%%%%%%%%%%%%%%%%%%%%%%%%%%%%%%%%%%%%%%%%%%%%%%%%%%%%
\begin{figure}[htb]
	\centering
	\includegraphics[width=\columnwidth]{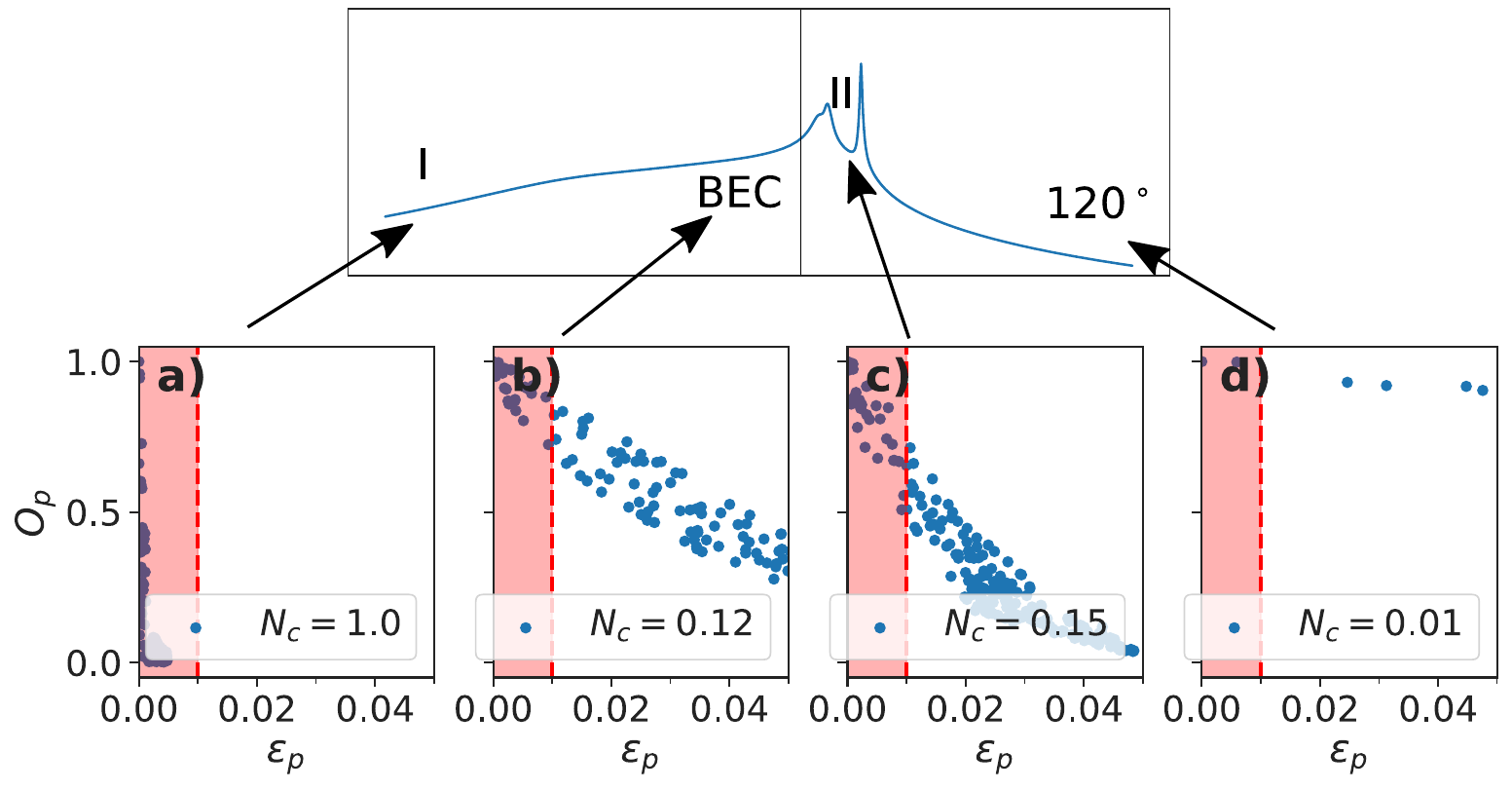}
	\caption{Results of the RTBC calculation for (a) $g=-5$, (b) $g=0$, (c) $g=0.73$ and (d) $g=4.0$. As detailed in the text, $N_{c}\approx 1$ signals a disordered phase, where $N_{c}\approx 0$ is expected for an ordered state. Consequently, we identify phase I as a possible spin-liquid candidate regime, {\resubmark whereas regime II shows no distinct features.}}
	\label{fig:RTBC_scatter_plot}
\end{figure}
%%%%%%%%%%%%%%%%%%%%%%%%%%%%%%%%%%%%%%%%%%%%%%%%%%%%%%%%%%%%%
%
As the authors explain in \cite{CastellsGraells2019}, the distribution of $O_{p}$ over $\epsilon_{p}$ for a set of ground-state vectors (all relevant parameters of the Hamiltonian remaining the same) depends strongly on whether a quantum phase is ordered or disordered. For an ordered phase, there exists a definitive boundary condition which accommodates the order intrinsic to the ground-state of an infinite system, whereas all other boundary conditions prohibit it. An anti-ferromagnetic spin-1/2 chain in 1D is a simple example. Here, if the number of sites in the system is odd, the anti-ferromagnetic order is prevented if no twisted boundary conditions are in place. Including TBC, $\theta=\pi$ is uniquely suited to minimize the ground-state energy as it accommodates the order of the system. As $\theta$ is altered from its optimal value of $\pi$, the ground state energy increases. Therefore, we expect only very few ground-states of similar energy to the optimal ground-state energy for an ordered phase. In the disordered case however, many different boundary conditions lead to very similar ground-state energies, including states that have very little overlap with the optimal ground-state. To quantify this distribution we define the set of configurations with energy comparable to the optimal ground state
\begin{align}
Q = \left\{\left(\theta_{x}^{(p)},\theta_{y}^{(p)}\right)\in P: \epsilon_{p}<\alpha\right\},
\end{align} 
where $\alpha$ is chosen suficiently small, e.g. $\alpha=0.01$. Subsequently, we define the fraction of low energy configurations to be
\begin{align}
N_{c}=\frac{\abs{Q}}{M}.
\end{align}
Continuing the argument from before, $N_{c}$ is typically small for ordered phases and close to unity for disordered ones. This characterization is somewhat dependent on the particular choice of $\alpha$, which also depends on the system size $\systemsize$. For our case of $\systemsize=24$ (we use the shape 24C shown in Fig.~\ref{fig:TwoD_ShapeCluster}) the choice of $\alpha=0.01$ is reasonable (see \cite{CastellsGraells2019}).

In Fig.~\ref{fig:RTBC_scatter_plot} we show the results of the RTBC calculations for the shape 24C, having performed $M=200$ realizations of twisting angles for each value of $g$ that we consider. In Fig.~\ref{fig:RTBC_scatter_plot}d), not all points are visible since we limit the $\epsilon_{p}$-axis to 0.05. Judging by the values of $N_{c}$ for each $g$, we find a clearly disordered regime for ${g=-5}$ (regime $\mathrm{I}$ in Fig.~\ref{fig:GS_fidelity_shapes_full}), with a strongly ordered phase at ${g=4}$ ($\ang{120}$-order). The {\resubmark RTBC results 
at value $g=0.73$, i.e. in phase II, are  in between the clearly disorderd situation shown in
Fig.~\ref{fig:RTBC_scatter_plot} a) and the BEC phase, shown in Fig.~\ref{fig:RTBC_scatter_plot} b), and a clear identification as an ordered or disordered phase is not easy. 
We will show, however, in the following sub-section that the situation becomes much clearer if the competing effects from the repulsive density-density interaction are switched off.}

%%%%%%%%%%%%%%%%%%%%%%%%%%%%%%%%%%%%%%%%%
\subsection{{\resubmark Hamiltonian without density-density interaction}}
%%%%%%%%%%%%%%%%%%%%%%%%%%%%%%%%%%%%%%%%%
%
%%%%%%%%%%%%%%%%%%%%%%%%%%%%%%%%%%%%%%%%%%%%%%%%%%%%%%%%%%%%%
\begin{figure}[htb]
	\centering
    \includegraphics[width=0.95\columnwidth]{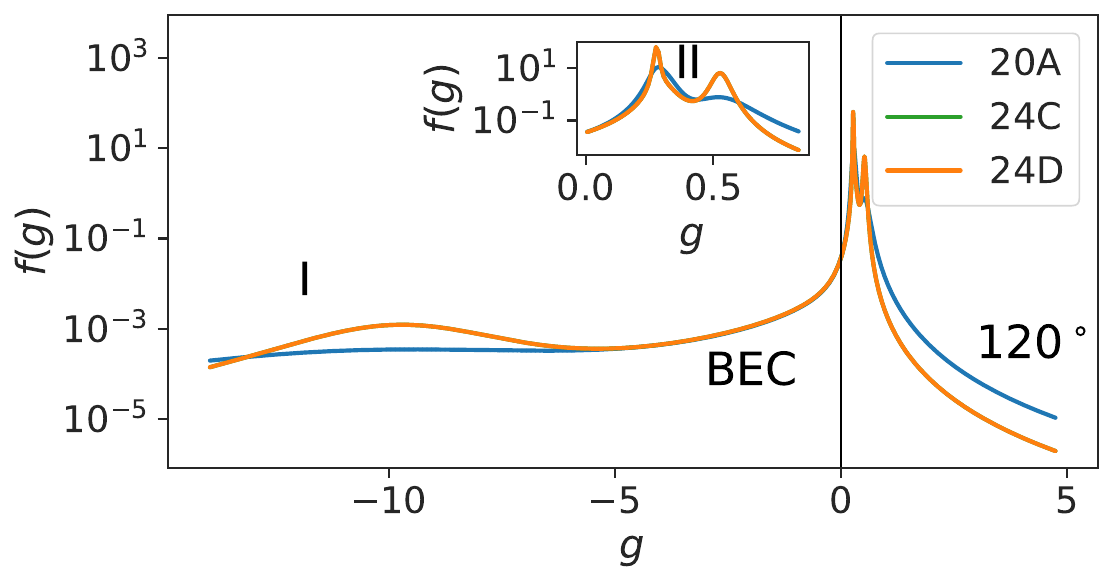}
	\caption{{\resubmark Ground-state fidelity metric $f$ as a function of the parameter $g$ for {the Hamiltonian \eqref{eqn:Rydberg_BHH_NoStark}}. The peaks of $f$ agree well for the different shapes and indicate potential phase transitions. The inset shows regime $\mathrm{II}$ in detail. We find the same qualitative behavior as for Hamiltonian \eqref{eqn:Rydberg_BHH_fullH}, but the transition points are shifted.}}
	\label{fig:GS_fidelity_shapes_NoStark}
\end{figure}
%%%%%%%%%%%%%%%%%%%%%%%%%%%%%%%%%%%%%%%%%%%%%%%%%%%%%%%%%%%%%
%
{\resubmark The character of phase II is somewhat masked by the simultaneous presence of a density-density interaction, which
drives the system into trivial ordered states. Therefore, we consider a modified version of Hamiltonian \eqref{eqn:Rydberg_BHH_fullH}, where we artificially switch off the density-density interaction:
\begin{align}\label{eqn:Rydberg_BHH_NoStark}
\hat{H}=
&
-J\sum_{\langle i,j\rangle}\hat{b}_{j}^{\dagger}\hat{b}_{i}
-2gJ\sum_{\langle\langle i,j\rangle\rangle}\hat{b}_{j}^{\dagger}\hat{b}_{i}\mathrm{e}^{\pm\frac{2\pi\mathrm{i}}{3}}(1-\hat{n}_{ij}).
\end{align}
%

%%%%%%%%%%%%%%%%%%%%%%%%%%%%%%%%%%%%%%%%%%%%%%%%%%%%%%%%%%%%%
\begin{figure}[htb]
	\centering
	\includegraphics[width=\columnwidth]{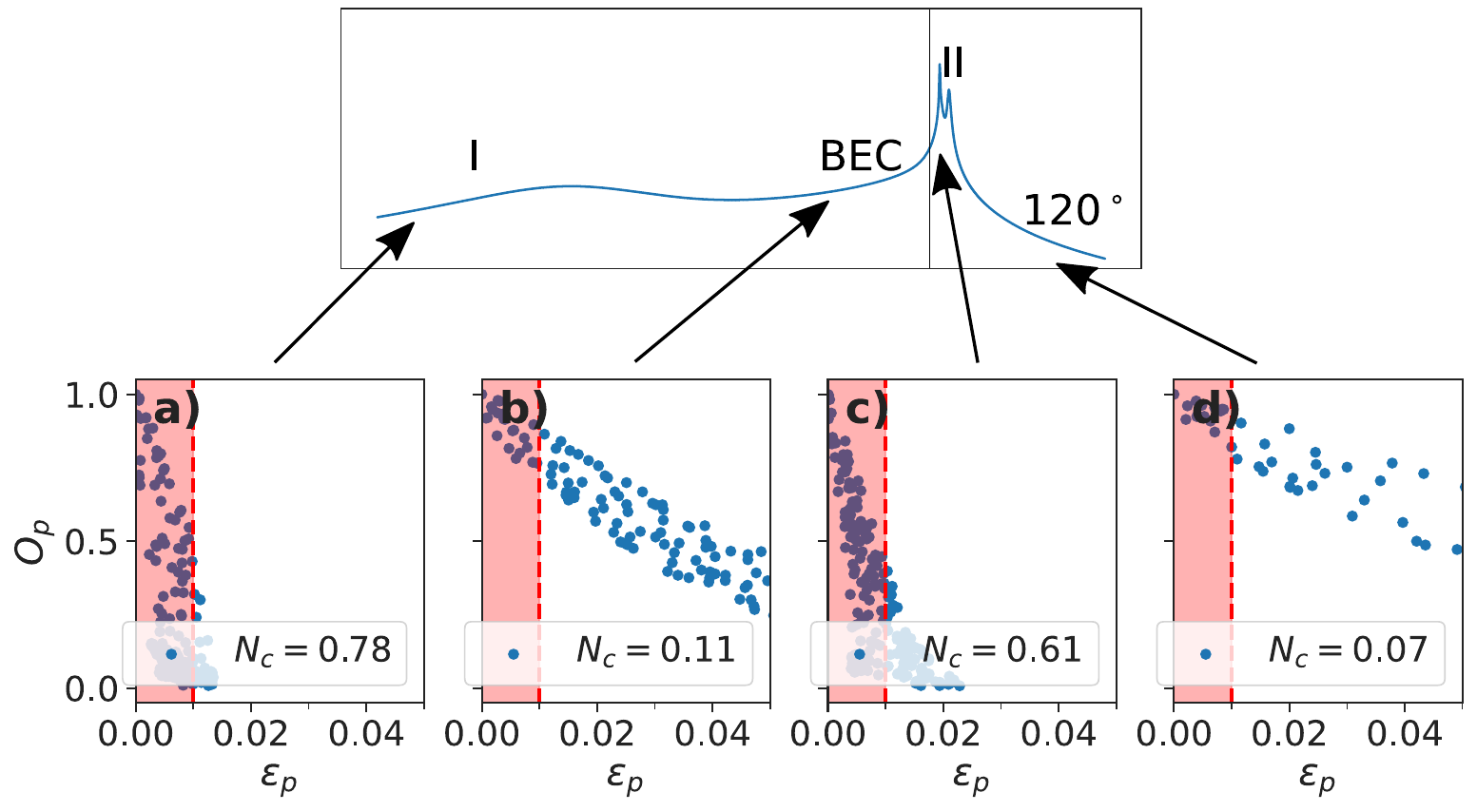}
	\caption{{\resubmark Results of the RTBC calculation for (a) $g=-15$, (b) $g=0$, (c) $g=0.42$ and (d) $g=4.0$ for Hamiltonian \eqref{eqn:Rydberg_BHH_NoStark}. As detailed in the text, $N_{c}\approx 1$ signals a disordered phase, where $N_{c}\approx 0$ is expected for an ordered state. We observe that regime II now shows strong disorder whereas for the full Hamiltonian we saw a distribution of ground states closer to the BEC case.}}
	\label{fig:RTBC_scatter_plot_NoStark}
\end{figure}
%%%%%%%%%%%%%%%%%%%%%%%%%%%%%%%%%%%%%%%%%%%%%%%%%%%%%%%%%%%%%

We first verify that the emergence of phases I and II is not due to density-density interactions. This can be seen from Fig.~\ref{fig:GS_fidelity_shapes_NoStark}, where we have plotted the ground-state fidelity metric for Hamiltonian 
\eqref{eqn:Rydberg_BHH_NoStark}.
In addition to the two peaks at small positive $g$ we find that the slow crossover to regime I at large negative $g$ is shifted to even larger absolute values.
Without the density-density interaction, the system is less likely to show long-range diagonal order. For this reason, we again consider the sensitivity to changes in boundary conditions (RTBC). The corresponding results are shown in Fig. \ref{fig:RTBC_scatter_plot_NoStark}. We observe that the intermediate phase II is now characterized by a large value of $N_c$, clearly signalling a disordered ground state, while
$N_c$ stays roughly the same for the other phases.}

%%%%%%%%%%%%%%%%%%%%%%%%%%%%%%%%%%%%%%%%%%
%%%%%%%%%%%%%%%%%%%%%%%%%%%%%%%%%%%%%%%%%%
\section{nature of the liquid phase} \label{sect:nature}
%%%%%%%%%%%%%%%%%%%%%%%%%%%%%%%%%%%%%%%%%%
%%%%%%%%%%%%%%%%%%%%%%%%%%%%%%%%%%%%%%%%%%

Having established the liquid-like behaviour of phase II 
we will investigate in the following the possible nature of this phase.

%%%%%%%%%%%%%%%%%%%%%%%%%%%%%%%%%%%%%%%
\subsection{Many-body Chern Number}
%%%%%%%%%%%%%%%%%%%%%%%%%%%%%%%%%%%%%%%

\begin{figure}
	\centering
		\includegraphics[width=0.9\columnwidth]{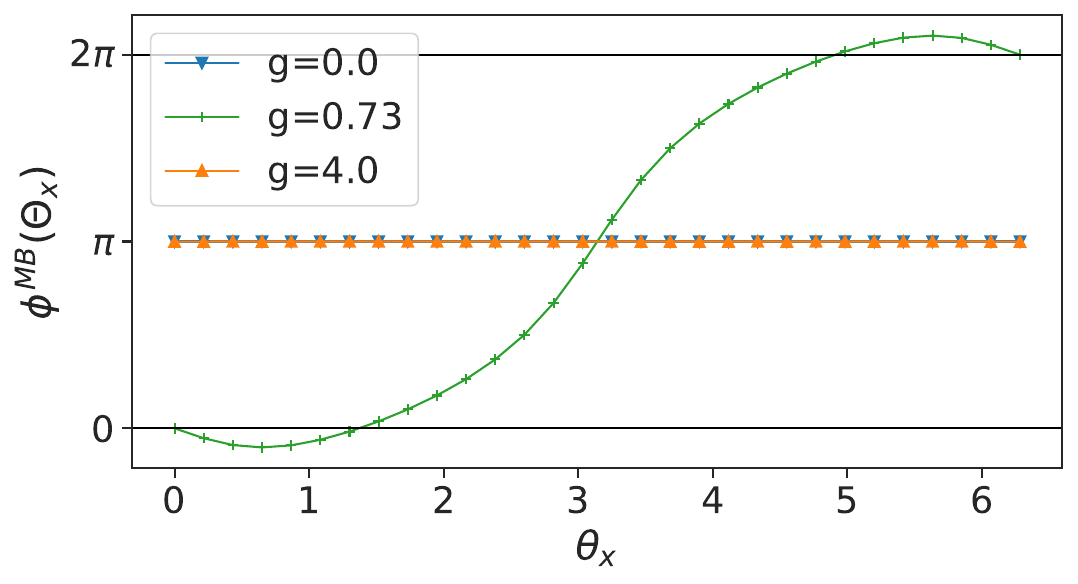}
	\caption{{\resubmark Many-Body Zak Phase winding for $g=0.73$ and the trivial cases for $g=0$ (BEC) and $g=4.0$ ($\ang{120}$, calculated for the full Hamiltonian \eqref{eqn:Rydberg_BHH_fullH}.}}
	\label{fig:Zak_winding_fullH}
\end{figure}

{\resubmark The complex NNN hopping in the Hamiltonian explicitly breaks time-reversal symmetry. We now show that as a consequence of this and of the nonlinear character of the NNN hopping, phase II is topological, characterized by a non-vanishing Chern number.}
The many-body Chern number in a two-dimensional lattice model on a torus can conveniently be obtained from the many-body ground state wavefunction 
$\ket{\Psi(\theta)}=\ket{\Psi\left(\theta^{x},\theta^{y}\right)}$ with twisted boundary conditions in $x$ and $y$ direction respectively:
\begin{equation}
    C = \frac{\mathrm{i}}{2\pi}\int_0^{2\pi}\!\!\! d\theta_x \int_0^{2\pi}\!\!\! d\theta_y \Bigl( \langle \partial_{\theta_x}\Psi(\mathbf{\theta})\vert \partial_{\theta_y}\Psi(\mathbf{\theta})\rangle - \textrm{c.c.}
    %\langle \partial_{\theta_y}\Psi(\mathbf{\theta})\vert \partial_{\theta_x}\Psi(\mathbf{\theta})\rangle
    \Bigr).
\end{equation}
For the numerical calculation we use a set of discrete twisting angles $\{\theta^{x}_{i},\theta^{y}_{j}\}$ where 
\begin{align}
    \theta^{\alpha}_{i} = \frac{2\pi}{D}i,
\end{align}
{$\alpha=x, y$} and $D$ is the number of intervals. We then calculate the ground-state wavefunction $\ket{\Psi\left(\theta^{x}_{i},\theta^{y}_{j}\right)}$ for each $\{\theta^{x}_{i},\theta^{y}_{j}\}$ and calculate the many-body Zak-phase using
\begin{align}
    \phi^{\textrm{MB}}\left(\theta^{y}_{j}\right)
    =& \textrm{Im}\ln \Bigl[
    \braket{\Psi\left(\theta^{x}_{1},\theta^{y}_{j}\right)}
    {\Psi\left(\theta^{x}_{2},\theta^{y}_{j}\right)}
    \\
    &\ldots
    \hspace{1cm}\braket{\Psi\left(\theta^{x}_{D},\theta^{y}_{j}\right)}
    {\Psi\left(\theta^{x}_{1},\theta^{y}_{j}\right)}\Bigr].\nonumber
\end{align}
Here, we take the loop product in the $x$-twisting angle only. The Chern number can then be calculated as the winding of the many-body Zak phase
\begin{align}
    C = \frac{1}{2\pi}\int_{0}^{2\pi}\!\!\! d\theta^{y}\, \pdv{\phi^{MB}}{\theta^{y}}.
\end{align}
In doing so, we obtain for ${g=0.73}$ a Chern number of $C=1$ to within numerical precision (see Fig. \ref{fig:Zak_winding_fullH}). We checked that the Chern number did not change when adding potential disorder of $\pm 0.1 J$. {\resubmark We also checked an additional shape (20A) and again obtain a Chern number of $C=1$}. For comparison, the calculation of the
Chern number in the $\ang{120}$ phase, i.e. for $g=4$, yields $C=0$ again to within numerical precision.

We did not calculate a Chern number for the disordered and gapless regime $\mathrm{I}$, as varying the twisting angles mixes the ground-state with excited states. 

%%%%%%%%%%%%%%%%%%%%%%%%%%%%%%%%%%%
%%%%%%%%%%%%%%%%%%%%%%%%%%%

%%%%%%%%%%%%%%%%%%%%%%%%%%%%%%%%%%%%%%%%%%
\subsection{Spin gap {\resubmark and collective gap}}
%%%%%%%%%%%%%%%%%%%%%%%%%%%%%%%%%%%%%%%%%%

In the study of disordered spin states, the distinction between gapless and gapped spin liquids is important. Therefore we now investigate the spin gap {\resubmark and the collective gap}.  

\begin{widetext}
%%%%%%%%%%%%%%%%%%%%%%%%%%%%%%%%%%%
\begin{figure*}[htb]
	\centering
		\includegraphics[width=1.8\columnwidth]{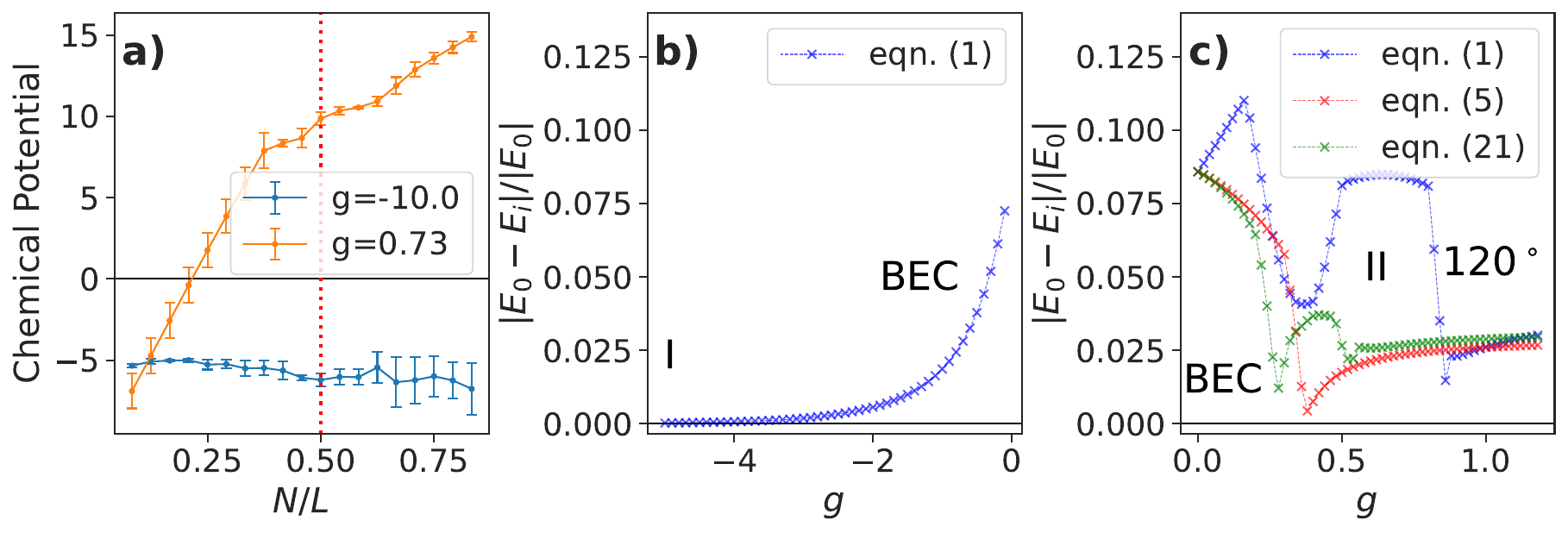}
	\caption{a) Chemical potential {for the full Hamiltonian \eqref{eqn:Rydberg_BHH_fullH}} on the shape $24C$  in the regimes $\mathrm{I}$ and $\mathrm{II}$, plotted over the density $\numparticles/\systemsize$, where $\numparticles$ represents the number of particles in the system. b) and c) {\resubmark Energy of the lowest excited state}, normalized with respect to the ground state, for cluster shape $24C$ and $\numparticles/\systemsize=1/2$ for regime $\mathrm{I}$ and $\mathrm{II}$. In phase $\mathrm{II}$ one recognizes a non-degenerate ground state. In phase $\mathrm{I}$ the energy gap is an order of magnitude smaller than in $\mathrm{II}$. The absolute energies per particle which we obtain at {\resubmark $g=-5$ are $E\approx -5.493\, |gJ|$, and at $g=0.73$ we find $E\approx -0.763\, |gJ|$.}}
	\label{fig:chemical_potential}
\end{figure*}
%%%%%%%%%%%%%%%%%%%%%%%%%%%%%%%%%%%
\end{widetext}

{\resubmark In a hard-core boson representation the spin gap corresponds to the change of the energy per particle when adding or subtracting a boson. 
Thus we calculate the chemical potential as the discrete first derivative of the many-body energy with regard to the particle number $\numparticles$.}
The result for the disordered regimes $\mathrm{I}$ and $\mathrm{II}$ can be seen in
Fig.~\ref{fig:chemical_potential}a. 
For the small systems that we are able to analyze numerically with exact diagonalization, the clear identification of a spin gap is masked by finite-size effects. Nevertheless while in phase I the spin gap clearly vanishes (see Fig.~\ref{fig:chemical_potential}a), 
the curve for phase II is indicative of a finite spin gap.

{\resubmark Secondly, we investigate the collective gap, which is shown  in Fig.~\ref{fig:chemical_potential}b) 
for the parameter regimes of the BEC and phase I
and in Fig.~\ref{fig:chemical_potential}c) for phase II. We plot the energy gap to the first excited state in the mean-field Hamiltomian (red symbols), the full Hamiltonian (blue symbols) and the 
Hamiltonian without density-density interaction terms (green symbols) for cluster shape 24C. We have also calculated the collective gap for few values of $g$ in clusters up to size 30 but could not make a reliable finite size scaling.
Phase I is clearly gapless, while for phase II we observe that the excitation gap is increased by the nonlinear hopping as compared to the mean-field case, and, as expected, the density-density interaction leads to an additional enhancement. This is indicative of a gapped phase II and thus a non-degenerate  ground state on a torus.}

The absence of degeneracy in the topological non-trivial regime would point to a symmetry protected topological phase (SPT). Our model has a $U\left(1\right)$ symmetry associated with particle-number conservation. There is a full classification of SPT phases \cite{Wen-PRB-2013} and for bosons in $d=2$ spatial dimensions with $U(1)$ symmetry different phases characterized by a $\mathbf{Z}$-quantized topological invariant exist, corresponding to the ${\cal H}^{1+d}[U(1),U(1)]$ cohomology group \cite{Wen-PRB-2013}.  
We note, however, that our finding of an odd-valued Chern number is different from the Chern numbers $C=\pm 2$ of the bosonic integer quantum Hall effect (BIQH) found e.g. in Ref.\cite{He2015} for bosons on a honeycomb lattice with NN and (different) density-dependent NNN hopping at unit filling and in Ref.\cite{Sterdyniak-PRL-2015} or \cite{Liu-PRB-2019} for bosons with internal degrees of freedom. 
{\resubmark The odd value of $C$ is also different than expected form the classification of interacting integer topological phases 
put forward in \cite{Lu-PRB-2012}.}

\begin{figure}[htb]
	\centering
		\includegraphics[width=0.98 \columnwidth]{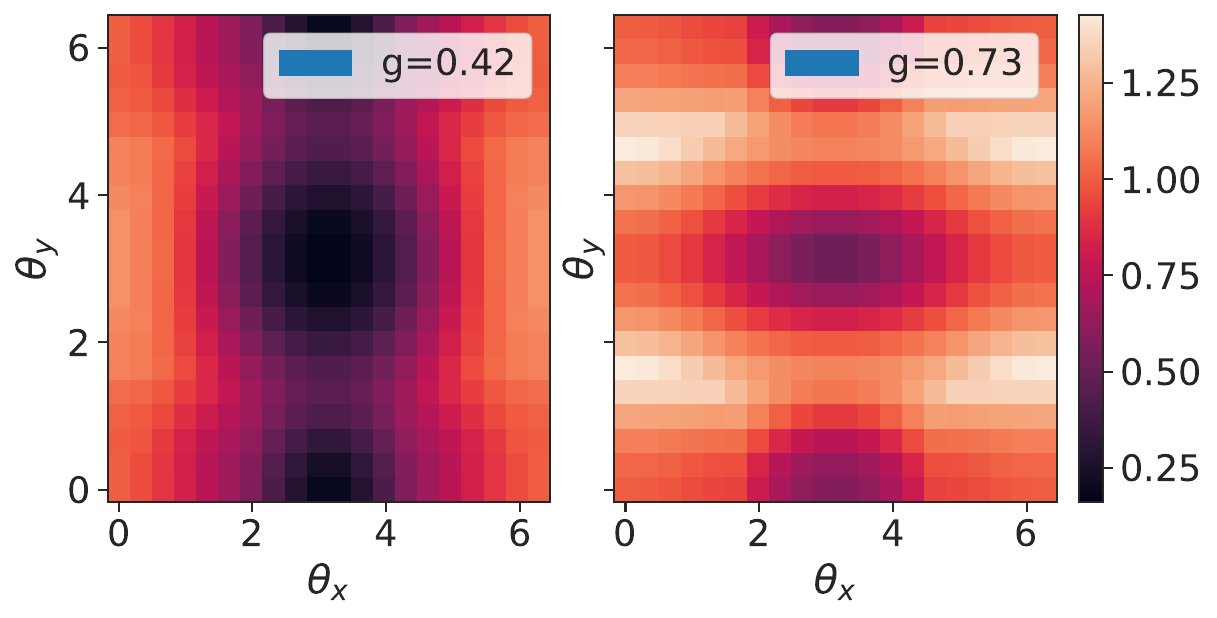}
	\caption{{\resubmark Collective gap as a function of twisted boundary angles $\theta_x$ and $\theta_y$ in phase II for shape $24C$, normalized to the gap at zero twist. \textit{left}: for Hamiltonian \eqref{eqn:Rydberg_BHH_NoStark}, i.e. without density-density interaction, and $g=0.42$. \textit{right}: for the full Hamiltonian \eqref{eqn:Rydberg_BHH_fullH} and $g=0.73$.}}
	\label{fig:gap_vs_twist}
\end{figure}

{\resubmark We are not able to perform a proper finite-size scaling using ED simulations and for the unambiguous verification of the gapfulness of phase II more sophisticated methods are needed. Corresponding investigations using DMRG and a novel tensor network approach are under way and will be reported elsewhere \cite{Marcello}. In particular for a Dirac QSL it is difficult to verify the gaplessness of the system, but the dependence of the collective gap on twisting angles is a signature of such a gapless spin liquid \cite{PhysRevX.7.031020}. In Fig.~\ref{fig:gap_vs_twist} we have therefore analyzed the size of the  gap for different twists of the boundary conditions. While the gap remains always finite, it shows a strong dependence.}

%%%%%%%%%%%%%%%%%%%%%%%%%%%%%%%%%%%%%%
%%%%%%%%%%%%%%%%%%%%%%%%%%%%%%%%%%%%%%%%

%%%%%%%%%%%%%%%%%%%%%%%%%%%%%%%%%%%%%%%%%%%%%%%%%%%%%%%%
\subsection{Fermion Hamiltonian and Chern-Simons gauge field}
%%%%%%%%%%%%%%%%%%%%%%%%%%%%%%%%%%%%%%%%%%%%%%%%%%%%%%%%

In the following we argue that the origin of the topological phase II can be understood from a representation of the model in terms of fermions.
A mapping from  hard-core bosons to spinless fermions can be achieved in one dimension by a Jordan-Wigner transformation  \cite{Jordan1928}. In two dimensions, this is accomplished via a Chern-Simons (CS) transformation, whose lattice version \cite{Sedrakyan2015} reads
\begin{align}
    \hat{b}_{i} = \hat{c}_{i}\, \mathrm{e}^{\mathrm{i}\sum_{j\neq i}\arg\left(z_{i}-z_{j}\right)\hat{n}_{j}},
\end{align}
where $\hat{b}_j$ are hard-core boson operators, and $\hat{c}_j$ fermion operators.
Here,  $z_{j}=x_{j}+\mathrm{i}y_{j}$ are the complex positions in the 2D lattice.
(Note that we use a different sign convention as in \cite{Sedrakyan2015}). When applying the CS transformation to our Hamiltonian we find
\begin{align}
\hat{H}=
&
-J\sum_{\langle i,j\rangle}\hat{c}_{j}^{\dagger}\hat{c}_{i} \mathrm{e}^{\mathrm{i}\hat{B}_{ji}}
-2gJ\sum_{\langle\langle i,j\rangle\rangle}\hat{c}_{j}^{\dagger}\hat{c}_{i}\mathrm{e}^{\pm\frac{2\pi\mathrm{i}}{3}} \mathrm{e}^{\mathrm{i}\hat{B}_{ji}}(1-\hat{n}_{ij})
\nonumber\\
&+2gJ\sum_{\langle i,j\rangle}\hat{n}_{i}\hat{n}_{j},
\end{align}
where a Chern-Simons gauge field
\begin{align}
    \hat{B}_{ji}=\sum_{l\neq j,i}\left[\arg(z_{i}-z_{l}) - \arg(z_{j}-z_{l})\right]\hat{n}_{l}
\end{align}
appears. It is instructive to decompose this field into a mean-field and a fluctuation
part $\hat B_{ji} = \langle \hat B_{ji}\rangle + \delta \hat B_{ji}$,
where $\delta \hat B_{ji} = \sum_{l\ne j,i} \delta\hat B_{ji}^{(l)}$
and
\begin{equation}
   \delta\hat B_{ji}^{(l)}= \left[\arg(z_{i}-z_{l}) - \arg(z_{j}-z_{l})\right] \bigl(\hat n_l -\langle \hat n_l\rangle\bigr).
\end{equation}
The mean-field term can easily be evaluated for an infinite hexagonal lattice  at half filling, where $\langle \hat n_j\rangle =0.5$. One finds that $\langle \hat B_{ji}\rangle$ is to good approximation a multiple of $2\pi$ for the NN hopping terms
 and may be disregarded, while for the NNN hoppings it just compensates the terms $\pm 2\pi/3$ to within a few percent. Thus the system can approximately be described by a Haldane model for fermions in the topological trivial regime of real-valued NNN hoppings which interact with a fluctuation Chern-Simons field and have a density dependent 
 NNN hopping. 
\begin{align}
\hat{H}\approx 
&
-J\sum_{\langle i,j\rangle}\hat{c}_{j}^{\dagger}\hat{c}_{i} \mathrm{e}^{\mathrm{i}\delta \hat{B}_{ji}}
-2gJ\sum_{\langle\langle i,j\rangle\rangle}\hat{c}_{j}^{\dagger}\hat{c}_{i} \mathrm{e}^{\mathrm{i}\delta \hat{B}_{ji}}(1-\hat{n}_{ij})
\nonumber\\
&+2gJ\sum_{\langle i,j\rangle}\hat{n}_{i}\hat{n}_{j}.
\end{align}
We have seen in Sect.~\ref{subsect:mean_field_approx} that within a mean-field approximation of the projector $(1-\hat n_{ij})\to (1-\overline{n})$, the topological phase II disappears. We thus conclude that here the fluctuation CS field $\delta \hat B_{ji}$ can most likely be neglected. In the full model, however, the projector $(1-\hat n_{ij})$ generates an additional mean-field contribution resulting from the site in between the next nearest neighbors $i$ and $j$. Denoting this site  here as $l=0$ we find
\begin{eqnarray}
    &&\exp\Bigl(\mathrm{i}\sum_{l\ne j,i} \delta \hat B_{ji}^{(l)}\Bigr)\,  (1-\hat n_0) = \\ &&\quad = \exp\Bigl(\mathrm{i}\!\!\sum_{l\ne j,i,0} \delta \hat B_{ji}^{(l)}\Bigr) \, \mathrm{e}^{\pm \frac{\pi \mathrm{i}}{3}}\,  (1-\hat n_0) \approx \mathrm{e}^{\pm \frac{\pi \mathrm{i}}{3}}\,  (1-\hat n_0).\nonumber
\end{eqnarray}
Assuming that all other contributions to the fluctuation CS field $\delta \hat B_{ji}^{(l)}$ are small and can be ignored, we recognize that the nonlinearity of the NNN hopping effectively generates a non-vanishing flux for the fermions which therefore enter a topologically non-trivial phase of the Haldane model with Chern number $C=1$. Thus we identify as the origin of the
topologically non-trivial phase II the additional Chern-Simons field created by the nonlinearity in the NNN hopping.

%%%%%%%%%%%%%%%%%%%%%%%%%%%%%%%%%%
\section{Experimental considerations: effects of longer-range interactions}\label{sect:long-range}
%%%%%%%%%%%%%%%%%%%%%%%%%%%%%%%%%%

\begin{figure}
	\centering
		\includegraphics[width=\columnwidth]{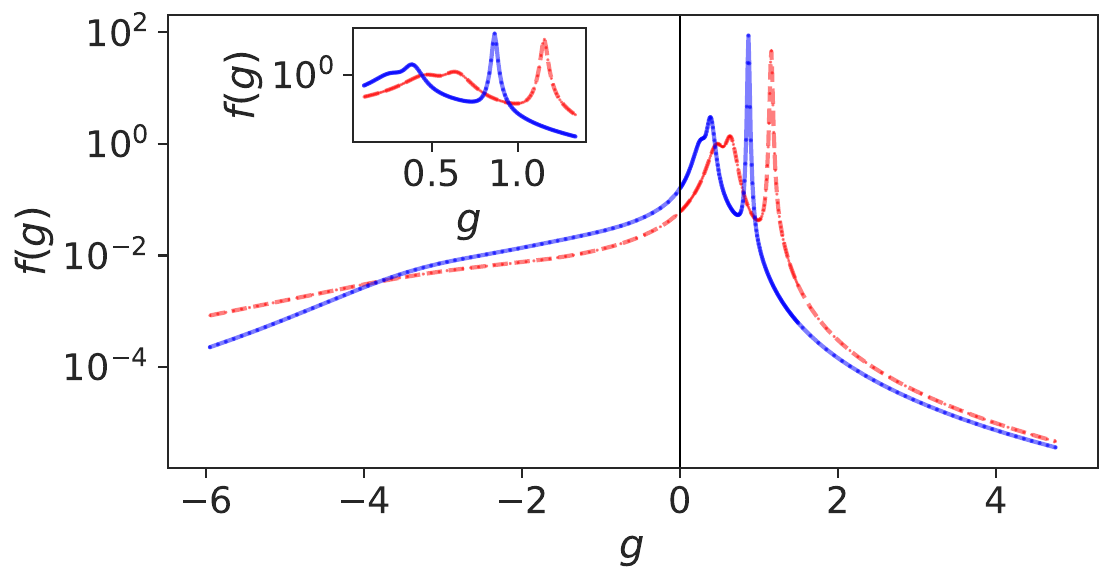}
	\caption{Comparison of ground state fidelity as function of $g$ for the cluster shape $24C$ with (red, dashed) and without (blue, solid) direct NNN hopping, resulting from expression \eqref{eqn:H-correction}.}
	\label{fig:fidelity_comparison}
\end{figure}

One of the main motivations of our work is to show that Rydberg excitations in an
array of trapped atoms are a suitable platform to observe spin liquids. We have
modelled the system with a simplified Hamiltonian \eqref{eqn:Rydberg_BHH_fullH} and thus some comments are in place about the limitations of this model.
The microscopic origin of the direct hopping of spin excitations is the resonant exchange of a micro-wave photon giving rise to a $\sim 1/r^3$ dipole-dipole coupling. In the Hamiltonian eq.\eqref{eqn:Rydberg_BHH_fullH} we have only taken into account direct  exchange couplings between nearest neighbors, neglecting the coupling to next nearest neighbors. These processes would lead to an
additional NNN hopping contribution to the Hamiltonian 
\begin{equation}
    \hat H_\textrm{LR}= - \frac{J}{(\sqrt{3})^3} \sum_{\langle\langle i,j\rangle\rangle} \hat b_j^\dagger \hat b_i,\label{eqn:H-correction}
\end{equation}
which does not affect phase I and the $\ang{120}$ phase for $\vert g\vert\to\infty$. It is however a sizeable modification of the NNN hopping in the regime of $g\lesssim 1$. We have numerically checked that the inclusion of \eqref{eqn:H-correction} does not compromise the emergence of the {non-trivial regime II} and only leads to a quantitative shift of the critical values for $g$. This is illustrated in Fig.~\ref{fig:fidelity_comparison} where we have compared the ground state fidelity with and without 
$\hat H_\textrm{LR}$. 

%%%%%%%%%%%%%%%%%%%%%%%%%%%%%%%%%%%
%%%%%%%%%%%%%%%%%%%%%%%%%%%

%%%%%%%%%%%%%%%%%%%%%%%%%%%%%%%%%%%%%%%%%%%%%%%%%%%%%%
\section{summary and conclusion} \label{sec:summary}
%%%%%%%%%%%%%%%%%%%%%%%%%%%%%%%%%%%%%%%%%%%%%%%%%%%%%%

Despite numerous experimental indications, the realization and verification of a quantum spin liquid phase in solid-state systems remains a major challenge. 
In the present paper we have proposed a model system accessible in cold-gas experiments with Rydberg atoms where
such a state could be realized and studied in extended parameter regimes.
Motivated by the recent experimental observation of nonlinear, complex hopping processes of Rydberg
spin excitations in two-dimensional arrays of trapped atoms, we
analyzed the many-body ground state of these excitations in a honeycomb lattice at half filling using exact diagonalization simulations. The density-dependent complex hopping as well as a nearest neighbor density-density interaction arise from second-order processes of excitations between two Rydberg levels, whose strength can be controlled by tuning the energy of a third, off-resonant Rydberg state. If the nonlinear hopping is treated in mean-field approximation, the model is equivalent to the Haldane model in the topologically nontrivial phase with additional nearest neighbor interactions. Since the elementary constituents are here spin-$1/2$ particles or, equivalently, hard core bosons rather than fermions, the mean-field model has however no topological ground state. It instead possesses only two trivial phases, a condensate (BEC) with a preferred occupation of modes
with wave vectors close to zero, as well as a $\ang{120}$ phase with spiral spin order and a remaining $SO(2)$ rotational symmetry. A phase transition between the two phases occurs when the direct hopping is of similar amplitude as the second-order, next nearest neighbor hopping. In the full model
an additional phase, denoted as phase II, emerges close to the mean-field critical point and there are indications of a transition or a crossover into another phase for very large second-order hoppings of opposite sign. The latter regime is however outside the range of validity of the effective many-body Hamiltonian for the 
Rydberg system. We verified the absence of simple spin {\resubmark or dimer} order in the new 
phases {\resubmark and considering randomly twisted boundary conditions we found strong evidence that both phases are disordered.
This becomes particularly evident if the density-density interaction, which competes with the hopping processes and drives the system towards density order is switched off.} Since the complex, nonlinear hopping breaks both time-reversal and chiral symmetry, the 
spins can have a non-vanishing spin chirality, which attains a very large value in both new phases.
We calculated the many-body Chern number and found a value $C=1$ to within numerical precision in phase II, which was also shown
to be robust against potential disorder, and $C=0$ in the BEC and $\ang{120}$ phases. {\resubmark Furthermore we calculated the spin and collective gaps using ED simulations on finite lattices with periodic boundary conditions. Since for the system sizes that can be reached with exact diagonalisation it is not possible to make reliable extrapolations to the thermodynamic limit, we cannot draw definite conclusions here. While the ED results clearly point to a gapless phase I, we found indications for a finite spin gap and a finite collective gap in phase II, which was shown to originate from the nonlinear hopping rather than from the density-density interaction. This would point to a symmetry protected topological phase protected by the $U(1)$ symmetry.} Considering a mapping of the hard-core boson Hamiltonian to spinless fermions coupled to a Chern-Simons field, we showed  that the topologically non-trivial phase is caused by the density dependence of the NNN hopping. The nonlinearity of this hopping generates an additional Chern-Simons flux for the fermion model which becomes topologically non-trivial due to this.
The odd value $C=1$ of the Chern number is however in contrast to several bosonic integer quantum Hall phases
found in Refs.\cite{Sterdyniak-PRL-2015,He2015,Liu-PRB-2019} and predicted from the general classification scheme of \cite{Lu-PRB-2012}. {\resubmark Calculating the collective gap in phase II for different twisted boundary conditions we found a nonzero, but
strongly varying value, which rather points to a gapless Dirac QSL, which is even more pronounced if the density-density interaction is switched off. Thus while there is strong evidence that phase II is a topological quantum spin liquid  at half filling,
its true nature remains unclear and requires further investigations.}

\emph{Note:} After finalizing this work we became aware of publication \cite{weber2022experimentally} predicting  a fractional quantum Hall phase for a similar system with average particle density $1/4$ 
induced by density-density interactions in engineered flat Chern bands.

%%%%%%%%%%%%%%%%%%%%%%%%%%%%%%%%%%%%%%%%%%%
\subsection*{Acknowledgement}
We thank Marcello Dalmonte, Poetri Tarabunga, Titas Chanda and Giuliano Giudici for very fruitful and inspiring discussions.
We furthermore thank Frank Pollmann and Tigran Sedrakyan for useful comments. The authors gratefully acknowledge financial support from the DFG through SFB TR 185, project number
277625399.

\subsection*{Author contributions}
M.F. conceived and supervised the project.
S.O. performed the analytic calculations and the numerical simulations.
Both S.O and M.K.-E. worked on the initial numerical implementation and all authors
analyzed the data. S.O. and M.F. worked on the final manuscript with
inputs of M.K.-E.

\bibliography{Rydberg_Haldane}

\end{document}